\documentclass[a4paper,11pt]{article}
\pdfoutput=1 

\usepackage{jcappub} 

\usepackage{multirow}
\usepackage{booktabs}
\usepackage{makecell} 
\usepackage{tabularx}
\usepackage{hyperref}
\usepackage{listings}
\usepackage{mathtools}
\usepackage{amsmath}
\usepackage{amsfonts}
\usepackage{amssymb}
\usepackage{epsfig}
\usepackage{graphicx}
\usepackage{bm}
\usepackage{array}
\usepackage{hyperref}
\usepackage{listings}
\usepackage{color}
\usepackage{float}
\usepackage[normalem]{ulem}
\usepackage{placeins}
\usepackage{verbatim}
\usepackage{enumerate}

\usepackage[T1]{fontenc} 

\usepackage{tikz,xcolor,hyperref}

\definecolor{lime}{HTML}{A6CE39}
\DeclareRobustCommand{\orcidicon}{\hspace{-1mm}
 \begin{tikzpicture}
 \draw[lime, fill=lime] (0,0) 
 circle [radius=0.16] 
 node[white] {{\fontfamily{qag}\selectfont \tiny \,ID}};
 \draw[white, fill=white] (-0.0525,0.095) 
 circle [radius=0.007];
 \end{tikzpicture}
 \hspace{-3mm}
}

\foreach \x in {A, ..., Z}{\expandafter\xdef\csname orcid\x\endcsname{\noexpand\href{https://orcid.org/\csname orcidauthor\x\endcsname}
 {\noexpand\orcidicon}}
}

\title{The Unknowns of the Diffuse Supernova Neutrino Background Hinder New Physics Searches}

\author[a,b]{Miller MacDonald\orcidA{},}
\author[a]{Pablo Mart{\'\i}nez-Mirav{\'e}\orcidB{},}
\author[a]{and Irene Tamborra\orcidC{}}

\affiliation[a]{Niels Bohr International Academy and DARK, Niels Bohr Institute, University of Copenhagen,
Blegdamsvej 17, 2100, Copenhagen, Denmark}
\affiliation[b]{Department of Physics $\&$ Laboratory for Particle Physics and Cosmology,
Harvard University, Cambridge, MA 02138, USA}

\emailAdd{mmacdonald@college.harvard.edu}
\emailAdd{pablo.mirave@nbi.ku.dk}
\emailAdd{tamborra@nbi.ku.dk}

\abstract{
Neutrinos traveling over cosmic distances are ideal probes of new physics. We leverage on the approaching detection of the diffuse supernova neutrino background (DSNB) to explore whether, if the DSNB showed departures from  theoretical predictions, we could  attribute such  modifications  to new physics unequivocally. In order to do so, we focus on  visible neutrino decay. 
Many of the signatures from neutrino decay are degenerate with  astrophysical unknowns entering the DSNB modeling. Next generation neutrino observatories, such as  Hyper-Kamiokande, JUNO, as well as DUNE, will  set stringent limits on   a neutrino lifetime over mass ratio $\tau/m \sim 10^{9}$--$10^{10}$~s~eV$^{-1}$~at $90\%$ C.L.,  if  astrophysical uncertainties and detector backgrounds were to be  fully under control.
However, if the lightest neutrino is almost massless and the neutrino mass ordering is normal, constraining visible decay will not be realistically possible in the coming few decades. We also assess the challenges of distinguishing among different new physics scenarios (such as visible decay, invisible decay, and quasi-Dirac neutrinos), all leading up to similar signatures in the DSNB. 
This work shows that the DSNB potential for probing new physics  strongly depends on an improved understanding of the experimental  backgrounds at next generation neutrino observatories as well as progress in the DSNB  modeling. 
}

\begin{document}
\maketitle

\section{Introduction}
\label{sec:intro}
When addressing the origin of non-zero neutrino masses, it is frequent to invoke extensions of the Standard Model in which neutrinos acquire non-standard properties. However,  data from oscillation and scattering experiments  show no conclusive hint of new physics in the neutrino sector to date~\cite{Huber:2022lpm,Giunti:2022aea,Arguelles:2022tki}. The quest for signatures of such non-standard neutrino properties in astrophysical sources and cosmology has so far met the same destiny~\cite{Gerbino:2022nvz,Arguelles:2022tki}.

The suite of next-generation multi-purpose neutrino observatories, currently  being developed or under construction, with larger fiducial volumes, improved detection technology, and refined background modeling,  will significantly enhance our neutrino detection rate~\cite{Huber:2022lpm},  shedding light on the eventual existence of new physics.  In turn, this would lead to a more precise measurement of the neutrino spectrum from all sources on Earth~\cite{Vitagliano:2019yzm}.

The diffuse supernova neutrino background (DSNB) is one of the most awaited discoveries in neutrino physics and astrophysics. It consists of the cumulative flux of neutrinos from all core-collapse supernovae exploding in our Universe, with energy of several MeV to a few tens of MeV~\cite{Vitagliano:2019yzm,Ando:2023fcc}. The recent loading of Gadolinium in the Super-Kamiokande water-Cherenkov tank has enhanced the detection prospects for this signal~\cite{Beacom:2003nk,Super-Kamiokande:2023xup,Super-Kamiokande:2024kcb, SK-Gd:2024}. Future neutrino experiments--JUNO~\cite{JUNO:2015zny,JUNO:2021vlw}, Hyper-Kamiokande~\cite{Hyper-Kamiokande:2018ofw,Hyper-Kamiokande:2016srs}, and DUNE~\cite{DUNE:2020ypp,Moller:2018kpn}--will also be sensitive to the DSNB  advancing our comprehension of the  core-collapse supernova population.

For a long time, the DSNB  has been considered an ideal probe of new physics in the neutrino sector~\cite{Ando:2003ie,Fogli:2004gy,Goldberg:2005yw, Baker:2006gm,Farzan:2014gza,Jeong:2018yts,DeGouvea:2020ang,Tabrizi:2020vmo,Ivanez-Ballesteros:2022szu,deGouvea:2022dtw,Das:2022xsz}, with supernovae being precious laboratories of fundamental physics, including neutrino physics~\cite{Mirizzi:2015eza,Horiuchi:2018ofe}. In particular, neutrinos from the DSNB travel over cosmological distances and can potentially  test the new physics scenarios whose signatures emerge over cosmic distances. 
Given the fascinating physics opportunities that would stem from the DSNB detection, it is timely to address the following question: \textit{If the DSNB showed departures from our theoretical predictions, could we unequivocally attribute the difference to Beyond the Standard Model (BSM) physics?} To answer this question, one needs to quantify the role of uncertainties of astrophysical origin in our modeling of the DSNB and account for them consistently. 

Let us assume our DSNB measurement was incompatible with predictions, even after accounting for the modeling uncertainties. Then, a second question arises: \textit{Could we discriminate between different BSM predictions and determine the physical origin of the unexpected signatures?} In some scenarios, one expects very distinctive features, such as dips or spikes in the spectrum~\cite{Goldberg:2005yw, Baker:2006gm,Farzan:2014gza,Jeong:2018yts,Balantekin:2023jlg}. Other models, however, predict soft spectral distortions and normalization features that are highly degenerate between them~\cite{Ando:2003ie,Fogli:2004gy,DeGouvea:2020ang,deGouvea:2022dtw,Tabrizi:2020vmo,Ivanez-Ballesteros:2022szu,deGouvea:2022dtw,Bell:2022ycf}. 

In this paper, we investigate how neutrino visible decay would modify the DSNB spectrum and present the projected sensitivity at future neutrino observatories, after accounting for the astrophysical uncertainties in the DSNB modeling. We also illustrate the interplay between astrophysical uncertainties and distortions in the DSNB attributable to BSM physics. To this end, in Sec.~\ref{sec:nudecaymodels}, we overview the physics of neutrino decay.  Section~\ref{sec:modelDSNBdecay} focuses on  how neutrino decay affects the DSNB. We present our findings in Sec.~\ref{sec:results}, quantifying the degeneracies between decay signatures, detection-related uncertainties, and the unknowns in the DSNB modeling. Section~\ref{sec:degeneracies} discusses how astrophysical uncertainties could mask or mimic visible neutrino decay signatures. Furthermore, we explore other BSM scenarios affecting the DSNB, such as  invisible decay and quasi-Dirac neutrinos, and focus on the potential to discriminate among them. Finally, in Sec.~\ref{sec:conclusion}, we critically discuss our findings. In addition, Appendix~\ref{sec:DSNBdetection} presents  our calculation of the event rates in next generation neutrino observatories. In Appendix~\ref{sec:statanalysis}, we discuss the details of the statistical analysis. Finally, Appendix~\ref{sec:DSNB3nudecay} is devoted to two-body visible neutrino decay in the three-neutrino framework.

\section{Two body visible neutrino decay}
\label{sec:nudecaymodels}

Finite neutrino lifetimes are a common prediction in extensions of the Standard Model accounting for non-zero neutrino masses~\cite{Bahcall:1972my,Shrock:1974nd,Petcov:1976ff,Marciano:1977wx,Zatsepin:1978iy,Chikashige:1980qk,Gelmini:1980re,Pal:1981rm,Schechter:1981cv,Shrock:1982sc,Gelmini:1983ea,Bahcall:1986gq,Nussinov:1987pc,Frieman:1987as,Kim:1990km}. Yet, for lifetimes longer than the age of the Universe, neutrinos would be effectively stable~\cite{Giunti:2014ixa}.
Depending on whether the decay products are experimentally accessible, decay channels are often categorized as \textit{visible} or \textit{invisible}. Throughout this work, we focus on the subset of visible neutrino decays in which there is at least one active neutrino in the final state. In this context, searches for  signatures of visible neutrino decay have been conducted using solar, atmospheric, accelerator, reactor, supernova, and high-energy neutrinos of astrophysical origin--see, e.g.,  Refs.~\cite{Lindner:2001th,Beacom:2002cb,Pagliaroli:2015rca,Gago:2017zzy,Funcke:2019grs,Abdullahi:2020rge,Bustamante:2020niz,Porto-Silva:2020gma,Picoreti:2021yct,deGouvea:2023jxn,Ivanez-Ballesteros:2023lqa}. 
These searches rely on the direct detection of ultra-relativistic neutrinos. 
Hence, in these searches the testable range of lifetimes is suppressed by the Lorentz factor. 
All these probes set limits on the ratio between neutrino mass and lifetime (i.e.~$m_i/\tau_i$) for the  mass eigenstates $i$. We find it convenient to report our findings also in terms of the product of the mass and the decay width in the neutrino rest frame, $\Gamma_i$, namely
\begin{align}
    \alpha_i = m_i\Gamma_i = \frac{m_i}{\tau_i} \, .
    \label{eq:alpha}
\end{align}

Additional limits on neutrino decay come from the cosmic microwave background~\cite{Basboll:2008fx,Archidiacono:2013dua,Hannestad:2004qu,Hannestad:2005ex,Escudero:2019gfk,Barenboim:2020vrr,Chacko:2019nej, FrancoAbellan:2021hdb,Chen:2022idm}  and Big Bang Nucleosynthesis~\cite{Escudero:2019gfk}. Those probes provide indirect (complementary) constraints on neutrino properties. Alternatively, a future detection of the cosmic neutrino background would also test whether neutrinos are stable~\cite{Akita:2021hqn,Huang:2024tbo}. Note that, contrary to all other probes, at least two of these cosmological neutrinos are non-relativistic. Therefore, the cosmic neutrino background would be sensitive to the neutrino lifetime, instead of the ratio $\tau_i/m_i$.

The relic supernova neutrinos, with energies in the MeV range and having traveled over cosmological distances, meet the ideal conditions to investigate neutrino decay. 
We consider that neutrinos are Majorana fermions and couple to a Majoron ($X$), or invisible massless Majoron-like particle~\cite{Kim:1990km}~\footnote{Although we focus on a specific neutrino decay model, we comment on the similarities with other scenarios and the applicability of our results in Sec.~\ref{sec:bsm}.}.
Such coupling would be responsible for the decay
\begin{align}
    \nu_j\to \nu_i \, + X \, ,
\end{align}
where $\nu_j$ and $\nu_i$ denote the initial- and final-state neutrino or antineutrino, respectively. The decay of $\nu_j$ to all lighter mass eigenstates is kinematically allowed and could be helicity flipping (h.f.) or helicity conserving (h.c.)~\footnote{Note that, since our discussion focuses on ultrarelativistic propagation, changes in helicity and chirality are  equivalent up to a factor $m_{i}/E_i$.}.
The total decay width of $\nu_j$ in the neutrino rest frame is 
\begin{equation}
    \Gamma_j = \sum_{m_i < m_j} \Big[\Gamma(\nu_j \to \nu_i) + \Gamma(\nu_j \to \bar{\nu_i})\Big]\, ;
\end{equation}
the corresponding branching ratios ($\mathtt{BR}$) are given by
\begin{equation}
    \label{eq:branchingratiodef}
    \mathtt{BR}(\nu_j \to \nu_i) = \frac{\Gamma(\nu_j \to \nu_i)}{\Gamma_j} \qquad \text{ and} \qquad \mathtt{BR}(\nu_j \to \bar{\nu_i}) = \frac{\Gamma(\nu_j \to \bar{\nu}_i)}{\Gamma_j}\, . 
\end{equation}

The width of each decay channel depends on neutrino masses~\cite{deGouvea:2022cmo}. The energy distribution of the final-state neutrino also depends on the energy of the initial-state one. We consider two extreme mass configurations that serve as reference for the varied phenomenology expected~\cite{deGouvea:2022cmo}: 
\begin{itemize}
    \item[-] When the mass difference between the initial and final-state neutrino is much larger than the final-state neutrino mass (i.e.,  $m_j - m_i \gg m_i$), we refer to the decays as being ``strongly hierarchical (SH).'' The energy distributions of the final-state neutrino are
\begin{equation}
    \label{eq:shenergyspectra}
    \psi_{\nu_j \to \nu_i,\,\mathrm{SH}}(E_j, E_i) = \frac{2E_i}{E_j^2} \qquad \text{and} \qquad \psi_{\nu_j \to \bar{\nu}_i,\,\mathrm{SH}}(E_j, E_i) = \frac{2}{E_j}\left(1 - \frac{E_i}{E_j}\right)\, ,
\end{equation}
for helicity-conserving and helicity-changing decays, respectively. The branching ratio of helicity-flipping and helicity-conserving decays are equal.

\item[-] Conversely, when the mass difference between both neutrinos is small compared to the absolute mass of the decay product (i.e.,  $m_j - m_i \ll m_i$), we denote these mass configurations as ``quasi-degenerate (QD).'' Helicity flipping decays are strongly suppressed and, therefore, neglected. The only relevant channels are the helicity-conserving ones; the energy distribution of the final-state neutrino is essentially the one of the initial-state neutrino:
\begin{equation}
\label{eq:qdenergyspectra}
    \psi_{\nu_j \to \nu_i,\,\mathrm{QD}}(E_j, E_i) = \delta(E_j - E_i) \, .
\end{equation}

\end{itemize}

Current data allow for two different mass orderings~\cite{deSalas:2020pgw,Esteban:2020cvm,Capozzi:2021fjo}: the  normal (NO) and inverted ordering (IO), where $m_{\nu_1} < m_{\nu_2} < m_{\nu_3}$ and $m_{\nu_3} < m_{\nu_1} < m_{\nu_2}$, respectively. Flavor oscillation experiments have determined the mass splittings, $\Delta m^2_{ji} = m^2_j - m^2_i$, accurately. Nonetheless, the absolute mass scale remains unknown~\footnote{Current bounds on the sum of neutrino masses assume neutrinos are stable. However, the constraints on neutrino masses and lifetimes from cosmology are closely intertwined. Hence, those limits do not apply straightforwardly. The next most constraining bound on the neutrino mass scale comes from measurements of the beta-decay spectrum, and the lightest neutrino could have a mass as large as $\sim 0.35$--$0.45$~eV~\cite{Katrin:2024tvg}.}. Thus, there are four possible extreme scenarios:
\begin{itemize}
    \item[-] {\it Normal ordering with strongly hierarchical masses} (NO SH), which corresponds to the lightest neutrino being (almost or completely) massless, $m_{\nu_1} \approx 0$. In this scenario,  all masses would be strongly hierarchical (i.e., $m_{\nu_1} \ll m_{\nu_2} \ll m_{\nu_3}$).
    \item[-] {\it Normal ordering with quasi-degenerate masses} (NO QD), resulting from large absolute mass scale. The three neutrino masses are much larger than the mass differences and $m_{\nu_1} \lesssim m_{\nu_2} \lesssim m_{\nu_3}$.
     \item[-] {\it Inverted ordering with a strong hierarchy between the lightest neutrino mass and the other two} (IO SH). If the lightest neutrino $\nu_3$ is close to massless, $m_{\nu_3} \approx 0$, then $m_{\nu_3} \ll m_{\nu_1} \lesssim m_{\nu_2}$. 
    \item[-] {\it Inverted ordering with quasi-degenerate masses} (IO QD). For large absolute mass scales, $m_{\nu_3} \lesssim m_{\nu_1} \lesssim m_{\nu_2}$.  We do not address this scenario; however, the phenomenology can be clearly understood from the same arguments that apply to other visible decay scenarios and, therefore, we comment on it when relevant.
\end{itemize}
When discussing the phenomenology of neutrino visible decays, we explore NO SH, NO QD, and IO SH. When relevant, we  comment on the scenario in which $m_{\nu_3}$ is large and the three masses are quasi-degenerate (i.e the IO QD scenario).

\begin{figure}
\centering 
\includegraphics[width=0.7\linewidth]{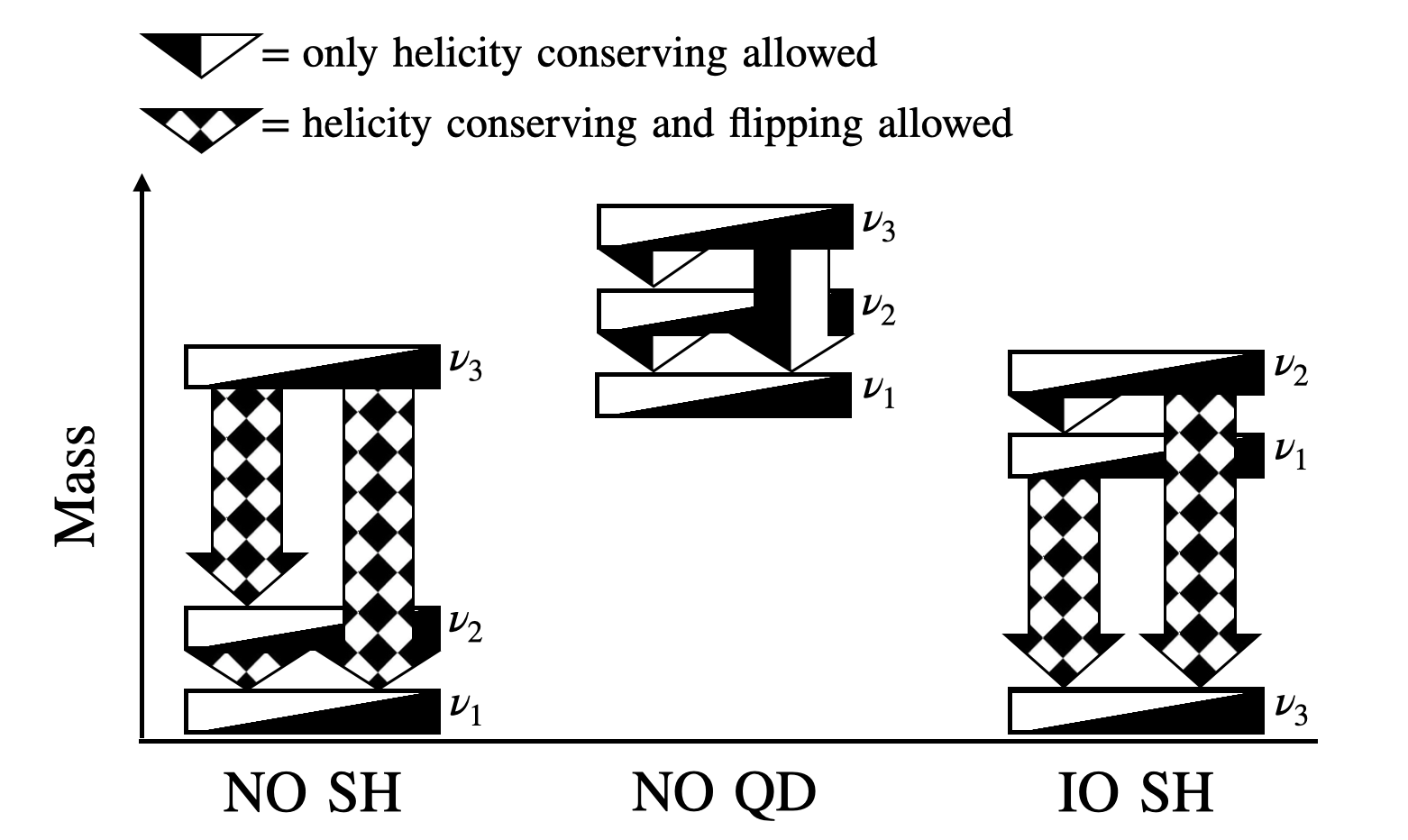}
\caption{\label{fig:decaychannelsfig} Allowed decay channels under the three mass configurations we consider: normal ordering with a strongly hierarchical absolute mass scale (NO SH, left), normal ordering with a quasi-degenerate absolute mass scale (NO QD, center), and inverted ordering with a strongly hierarchical mass scale (IO SH, right). Half-black half-white arrows denote that the allowed decays between the corresponding mass states must be helicity-conserving; checkered arrows indicate that the decays can be either helicity conserving or helicity flipping. The $y$-axis represents the absolute mass of the  mass states in each mass configuration. }
\end{figure}

Figure~\ref{fig:decaychannelsfig} shows the three different   mass configurations highlighted above (NO SH, NO QD, and IO SH). In each case, we indicate the allowed decay channels. Notice that, for quasi-degenerate mass pairs, only helicity-conserving decays can occur. On the contrary, for strongly hierarchical mass pairs, helicity flipping decays are also allowed––i.e., neutrinos decay into antineutrinos and vice versa.

\section{Diffuse supernova neutrino background in the presence of neutrino decay}
\label{sec:modelDSNBdecay}
In this section, we introduce the formalism adopted to model the DSNB signal in the presence of neutrino decay. We then explore the signatures induced on the DSNB by neutrino decay as well as the degeneracies existing among the different decay scenarios.
\subsection{General framework}
The DSNB depends on the supernova rate ($R_\mathrm{SN}$) and the time-integrated supernova neutrino yield ($F_{\nu_i}(E^0_\nu, M)$)  which are both functions of the redshift ($z$), the progenitor mass ($M$), and the emitted-neutrino energy ($E^0_\nu$).

In the presence of decays, the number density of relic supernova neutrinos is~\cite{Fogli:2004gy}
\begin{equation}
\label{eq:dsnb_with_decay}
\begin{aligned}
    n_{\nu_i}(E'_\nu, z) &= \int_{8M_\odot}^{M_\mathrm{max}}dM~\frac{1}{1+z}\int_z^{z_\mathrm{max}} \frac{dz'}{H(z')} \Biggr\{ R_\mathrm{SN}(z')F_{\nu_i}\left(E'_\nu\frac{1+z'}{1+z}, M\right) 
 \\ &+ \sum_{m_j > m_i}q_{ji}\left(E'_\nu\frac{1+z'}{1+z}, z'\right) \Biggl\}e^{-\alpha_i [\Upsilon(z') - \Upsilon(z)](1+z)/E'_\nu} \, ,
\end{aligned}
\end{equation}
where $E'_\nu$ is the observed neutrino energy at redshift $z$, and is related to the neutrino energy at redshift $z=0$ ($E_\nu$) as $E_\nu = E'_\nu/(1+z)$ The subscripts $i$ and $j$ denote the mass eigenstates, and the parameter $\alpha_i$ parametrizes the decay; see Eq.~\eqref{eq:alpha}. 
We consider supernova masses up to $M_\mathrm{max} = 125 M_\odot$ and neglect any contribution from $z_\mathrm{max} \gtrsim 5$~\cite{Ando:2004hc}. For the considered redshift range, we parameterize the expansion rate of the Universe as $H(z) = H_0\sqrt{\Omega_M(1+z)^3 + \Omega_\Lambda}$ and adopt  $H_0 = 70~\mathrm{km~s^{-1}~Mpc^{-1}}$~\cite{ParticleDataGroup:2022pth}, $\Omega_M = 0.3$ as well as $\Omega_\Lambda = 0.7$~\cite{Planck:2018vyg} for the local expansion rate, the energy density of dark energy and  of matter, respectively.

The terms $q_{ji}(E,z)$ in Eq.~\eqref{eq:dsnb_with_decay}-- where we have addopted the short-hand notation $E = E'_\nu (1+z')/(1+z)$--denote contributions to the neutrino number density of  a given mass eigenstate ($n_{\nu_i}$), from heavier mass states that have decayed at larger redshifts. Note that these terms would thus be absent for $n_{\nu_3}$ ($n_{\nu_1}$) in NO (IO). They can be written as
\begin{equation}
    q_{ji}(E, z') = \int_{E}^\infty d\tilde{E}_\nu~n_{\nu_j}(\tilde{E}_\nu, z') \frac{\alpha_j \mathtt{BR}(\nu_j \to \nu_i)}{\tilde{E}_\nu}\psi_{\nu_j \to \nu_i}(\tilde{E}_\nu, E)\ ,
\end{equation}
where $\psi_{\nu_j \to \nu_i}$ are the energy spectra defined in Eqs.~\eqref{eq:shenergyspectra} and~\eqref{eq:qdenergyspectra}. Finally, the exponential term in the integrand depends on the redshift-dependent function
\begin{equation}
    \label{eq:effectivelength}
    \Upsilon(z) = \int_0^z \frac{dz'}{H(z')}\frac{1}{(1+z')^2}\ ,
\end{equation}
and accounts for decays from $\nu_i$ into lighter mass states. It would thus be trivial for $n_{\nu_1}$ ($n_{\nu_3}$) in NO (IO).

We model the DSNB following Refs.~\cite{Moller:2018kpn,Martinez-Mirave:2024hfd}. This procedure considers a supernova rate normalization such that $\int_{8M_\odot}^{125M_\odot} dM~R_\mathrm{SN}(0, M) = (1.25 \pm 0.5) \times 10^{-4}\mathrm{~Mpc^{-3}~yr^{-1}}\, $\cite{Lien:2010yb}, which is one of the main theoretical uncertainties in the prediction of the DSNB. We compute the neutrino spectrum based on one-dimensional spherically symmetric hydrodynamical simulations~\cite{Mirizzi:2015eza,Garching_2015}, as in Ref.~\cite{Martinez-Mirave:2024hfd}. Note that our understanding of the impact of neutrino-neutrino self-interaction on the final neutrino spectra is incomplete~\cite{Tamborra:2020cul,Volpe:2023met}; this phenomenon is an additional   source of uncertainty, however likely resulting   in a $10\%$ change in the time-integrated flux~\cite{Lunardini:2012ne}. We split the neutrino population in two: black hole forming collapses and successful explosions; we  denote the fraction of black hole forming collapses with $f_\text{BH}$. Note that $f_\text{BH}$ could be degenerate with the fraction of magnetorotational collapses as shown in Ref.~\cite{Martinez-Mirave:2024zck}, but we neglect this contribution  for simplicity. For what concerns neutrino flavor conversion, we account for Mikheev-Smirnov-Wolfenstein (MSW) flavor conversion in the source~\cite{Dighe:1999bi,Lunardini:2012ne} and take into account  that neutrinos arrive at Earth as an incoherent sum of mass eigenstates, as in Ref.~\cite{Martinez-Mirave:2024hfd}. Finally, because DSNB neutrinos are ultrarelativistic, the flux at redshift $z=0$ of each mass eigenstate is simply related to the number density as $\phi_{\nu_i} = c n_{\nu_i}$, where $c$ denotes the speed of light.

\subsection{Signatures of two-body neutrino decays on the diffuse supernova neutrino background}
\label{sec:DSNB2nudecay}

We now investigate how neutrino decay affects the DSNB signal. For the decay of the mass state $\nu_i$ to begin yielding non-trivial phenomenology at Earth, its lab-frame lifetime must be approximately equal to its distance travelled ($L$):
\begin{equation}
    \label{eq:decay_pheno_condition}
    \frac{\tau_i}{m_i}  \lesssim \frac{L}{E_\nu}\qquad \text{or equivalently} \qquad \alpha_i  \lesssim \frac{E_\nu}{L}\, .
\end{equation}

If $\alpha_i$ is large enough (i.e.~$\alpha_i \gtrsim 10^{-24}~\mathrm{eV^2}$), almost all $\nu_i$ mass states  decay en route to Earth for characteristic DNSB detectable energies, $E_\nu \sim 10 - 30$~ MeV. This range of the parameter space corresponds to what we denote as {\it complete decay}. 
Decay parameters $\alpha_i \lesssim 10^{-26}~\mathrm{eV^2}$ correspond to characteristic decay lengths larger than the observable Universe ($E_\nu/\alpha_i \gtrsim 1/H_0$). Thus, the DSNB flux is almost unchanged for such small decay parameters.
For the intermediate range, i.e.~$10^{-26}~\mathrm{eV^2} \lesssim \alpha_i \lesssim 10^{-24}~\mathrm{eV^2}$, only a fraction of the neutrinos decay. Observable signatures are more prominent at the low energy end of the spectrum. We refer to this range of parameter space as {\it partial decay}.

Since the best detection prospects for the DSNB are in electron antineutrinos \cite{Super-Kamiokande:2021jaq, Hyper-Kamiokande:2018ofw, JUNO:2015zny}, we focus on how decays impact the $\bar{\nu}_e$ flux. However, for completeness, our analysis also considers the impact of neutrino decays on the detection prospects for the $\nu_e$ component~\cite{DUNE:2020ypp}. Note that the fraction of $\bar{\nu}_1$ in $\bar{\nu}_e$ is larger than the one of $\bar{\nu}_2$, and the fraction of $\bar{\nu}_3$ in $\bar{\nu}_e$ is almost negligible. Hence, decays that populate the mass eigenstate $\bar{\nu}_1$ are more likely to be observed, whereas decays to $\bar{\nu}_3$ would hamper the DSNB detection. This fact is a consequence of the size of the elements in the lepton mixing matrix, i.e. $|U_{e1}|^2 > |U_{e2}|^2 \gg |U_{e3}|^2$~\cite{deSalas:2020pgw}. 

Decay signatures in the DSNB are also heavily dependent on the exact decay channels that are allowed, their relative strength,  the absolute mass scale, and the mass ordering. If the channel is strongly hierarchical, the energy spectra result in decay products with less energy than the initial ones. This implies  more decay products being present at low energies, distorting the spectrum of the DSNB. If the channel is quasi-degenerate, all the parent energy is transferred to the daughter, and  the primary signature is a change of the DSNB normalization in the complete decay regime, without any spectral distortion. 

\begin{figure}[tbp]
\centering
\includegraphics[width=0.95\linewidth]{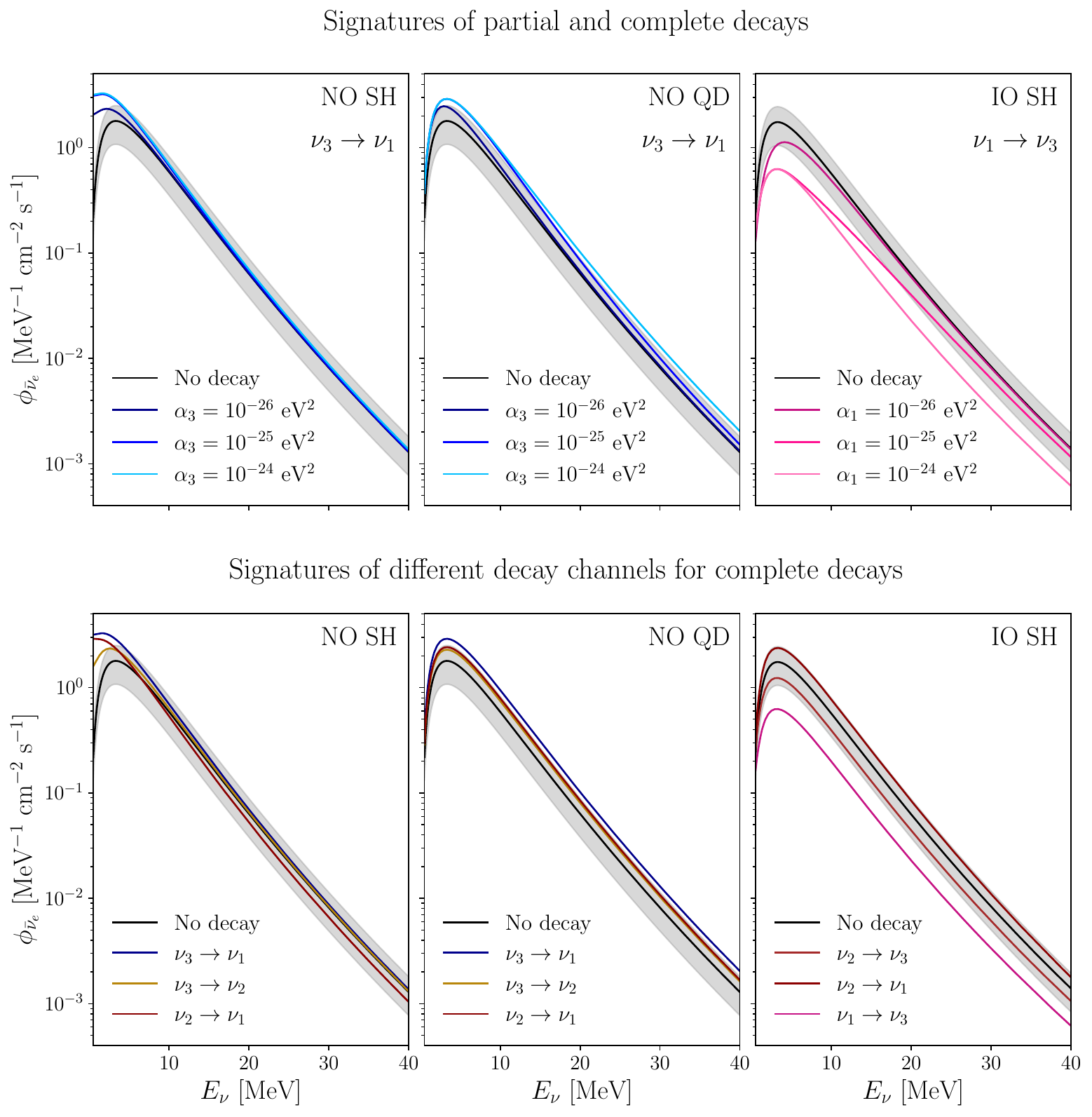}
\caption{\label{fig:DSNBfluxepartialcompletedecay} Summary of  the effects of neutrino decay in the  two-neutrino framework on the $\bar{\nu}_e$ component of the DSNB spectrum, for the partial and full decay scenarios.  
The $\bar\nu_e$ DSNB flux with no decay is plotted in black, and the gray band corresponds to the supernova rate uncertainty. \textit{Top panels:} Partial and complete decay scenarios. 
The left and central panels show the DSNB for different values of the  decay parameter, $\alpha_3$, for the decay channel $\nu_3 \rightarrow  \nu_1$, in the strongly hierarchical and quasi-degenerate limits. The right panel shows the DSNB flux in IO SH with for the  $\nu_1 \to \nu_3$ channel. Smaller decay parameter values (i.e.~longer lifetimes) are plotted with darker shades.  
\textit{Bottom panels:} 
Complete decay scenario with $\alpha = 10^{-24}~\mathrm{eV^2}$. The left and middle panels show the DSNB for NO SH and NO QD, respectively, for three decay channels: $\nu_3 \to \nu_1$ (blue), $\nu_3 \to \nu_2$ (orange), and $\nu_2 \to \nu_1$ (red). The right panel shows the DSNB for three IO SH decay channels: $\nu_2 \to \nu_3$ (brown), $\nu_2 \to \nu_1$ (red), and $\nu_1 \to \nu_3$ (magenta).}
\end{figure}

We show the DSNB in the presence of partial and complete decay for two different  decay channels in the top three panels of Fig.~\ref{fig:DSNBfluxepartialcompletedecay}. The three bottom panels display how different complete decay scenarios impact the DSNB. In this figure three main generic features can be observed: 
\begin{enumerate}[(i)]
    \item for partial decays observable signatures arise only at low energies, whereas for complete decays the whole spectrum is affected;
    \item depending on the mass eigenstate in the final state, the observable flux can be enhanced or damped; and 
    \item QD decays barely distort the spectral shape while strongly hierarchical ones shift neutrino energies to the lower end of the spectrum.
\end{enumerate}

Hereafter, we assume that only one decay channel is allowed at a time. This simplified approach--which is often referred to as \textit{effective 2-neutrino framework}--provides a comprehensive insight to the phenomenological implications of neutrino decay.

For the NO, decays from $\nu_3$ to either $\nu_2$ or $\nu_1$  increase the integrated electron antineutrino flux in the energy range above $10$~MeV, due to the very small value of $|U_{e3}|^2$. The $\nu_2 \to \nu_1$ decay channel would seem to also increase the flux, since $|U_{e1}|^2 > |U_{e2}|^2$. This is indeed the case if the channel is QD, where the energy distribution of initial and final state neutrinos is the same. However, if the channel is SH, the final DSNB spectrum is shifted to lower energies,  piling up  below $10$~MeV due to the abundance of $\nu_1$ (the decrease in $\nu_2$  at larger energies results in a flux \textit{decrease} at these energies). All these features can be observed in the bottom-left and bottom-central panels of Fig.~\ref{fig:DSNBfluxepartialcompletedecay}, for  complete decay.

For IO, decays to $\nu_3$ reduce the flux. For very light $\nu_3$, $\nu_2 \to \nu_3$ and $\nu_1 \to \nu_3$ decays are strongly hierarchical. However, since the decay product is $\nu_3$, the expected spectral distortions are not observable. Hence, the main signal is a normalization decrease. Conversely, $\nu_2 \to \nu_1$ decay is QD and results in a normalization increase, as in the case of normal ordering.

For inverted mass ordering and quasi-degenerate masses, i.e. for IO QD, the phenomenology of the decay $\nu_2 \to \nu_1$ is identical to the IO SH one. The observational differences in the spectrum from the decay $\nu_1 \to \nu_3$, i.e. the differences between IO SH and IO QD, are highly indistinguishable. The smallness of $|U_{e3}|^2$ would lead to almost identical observed $\bar{\nu}_e$ spectra, despite the different $\nu_3$ spectra.

\section{Future sensitivity to visible neutrino decay}
\label{sec:results}

In this section, we present our forecast on the constraining power of next-generation neutrino observatories––Hyper-Kamiokande, JUNO, and DUNE––to neutrino decay. Details on the technical characteristics of the neutrino observatories are provided in Appendix~\ref{sec:DSNBdetection}. 

Figure~\ref{fig:dsnbeventrates} shows the spectral distribution of the number of events expected at Hyper-Kamiokande, JUNO and DUNE after $20$~years of exposure time, energy bins of $2$~MeV, and  several decay hypotheses. For comparison, we also display the total background events for the same energy binning and exposure time.

\begin{figure}[tbp]
\centering 
\includegraphics[width=\linewidth]{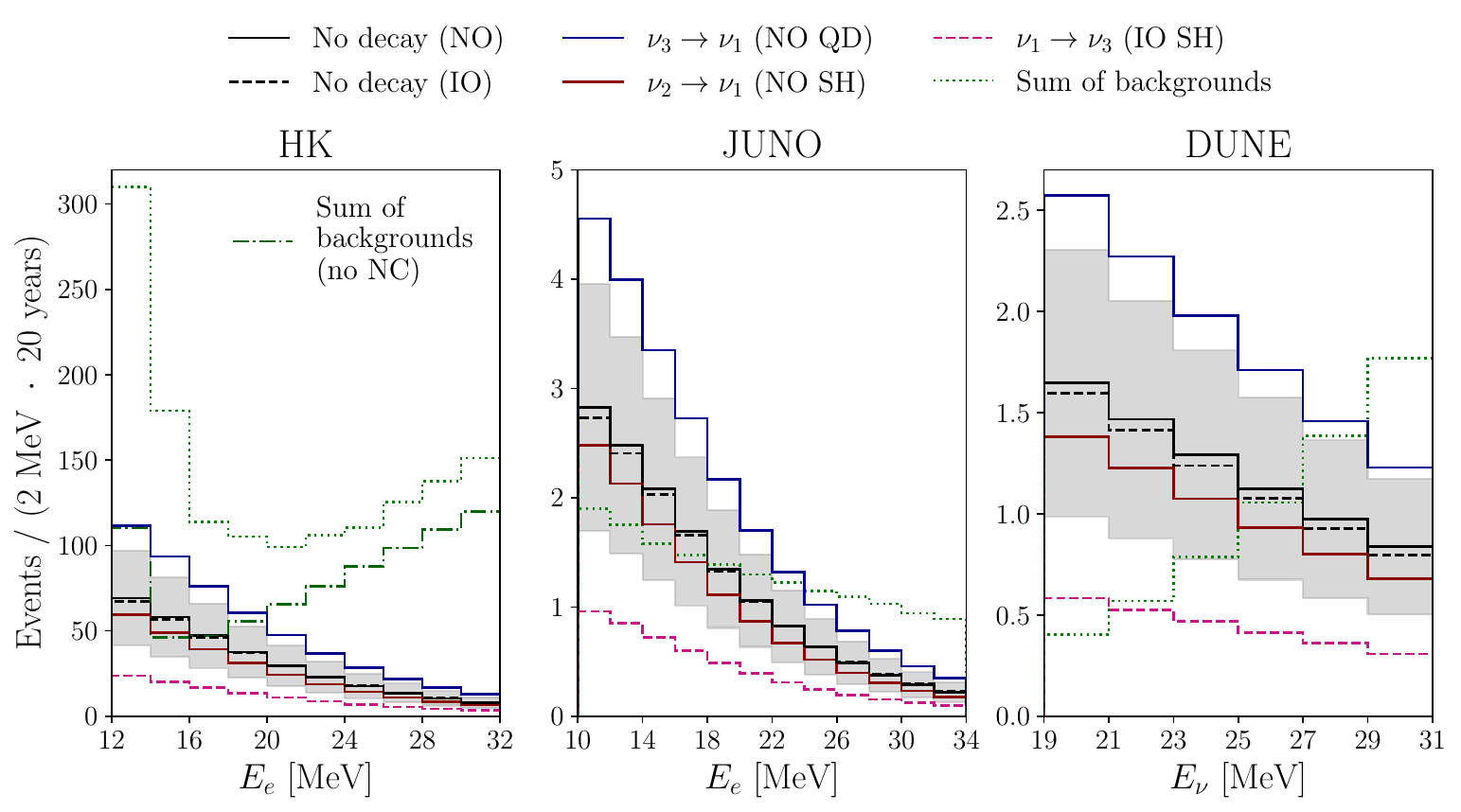}
\caption{\label{fig:dsnbeventrates} Event rates for $\alpha = 10^{-24}$~eV$^2$ for $20$~years of exposure as a function of the  reconstructed positron energy at Hyper-Kamiokande and JUNO in the left and central panels, respectively, and as a function of the reconstructed neutrino energy at DUNE in the right panel. Solid lines correspond to NO scenarios, while dashed lines correspond to the IO scenarios. Black lines correspond to the no decay hypothesis. Colored lines illustrate  complete effective $2\nu$ decay scenarios: $\nu_2 \to \nu_1$ NO SH (brown), $\nu_3 \to \nu_1$ NO QD (blue), and $\nu_1 \to \nu_3$ IO SH (magenta). The gray band represents the supernova rate uncertainty. Dotted light-green lines correspond to the sum of the expected background events. In the left panel, the dark green dash-dotted line represents the expected background events without the atmospheric neutral current events at Hyper-Kamiokande.}
\end{figure}

We calculate the combined sensitivity over $20$~years of Hyper-Kamiokande (HK), JUNO, and DUNE to various $2\nu$ decay channels under two different detection scenarios. In the  \textit{conservative} scenario, we assume current knowledge of the astrophysical uncertainties (i.e., the supernova rate and $f_{\rm BH}$) as well as detector backgrounds as  outlined in Appendix~\ref{sec:DSNBdetection}. In the  \textit{optimistic} scenario, we consider   significant progress in both reducing astrophysical uncertainties and extant backgrounds: the supernova rate normalization and the background rates are known to within a $5\%$ error of its current best fit values, all neutral-current atmospheric backgrounds in Hyper-Kamiokande are efficiently tagged, and the BH fraction is known (we fix  $f_\mathrm{BH} = 0.21$). Further sources of uncertainties exist in the modeling of the DSNB in both scenarios, cf., e.g., Refs.~\cite{Lunardini:2012ne,Kresse:2020nto,Nakazato:2015rya,Ashida:2023heb,Horiuchi:2020jnc,Martinez-Mirave:2024zck}. However, the uncertainty on the supernova rate should be the dominant one~\cite{Kresse:2020nto}.

\begin{figure}[tbp]
\centering 
\includegraphics[width=\linewidth]{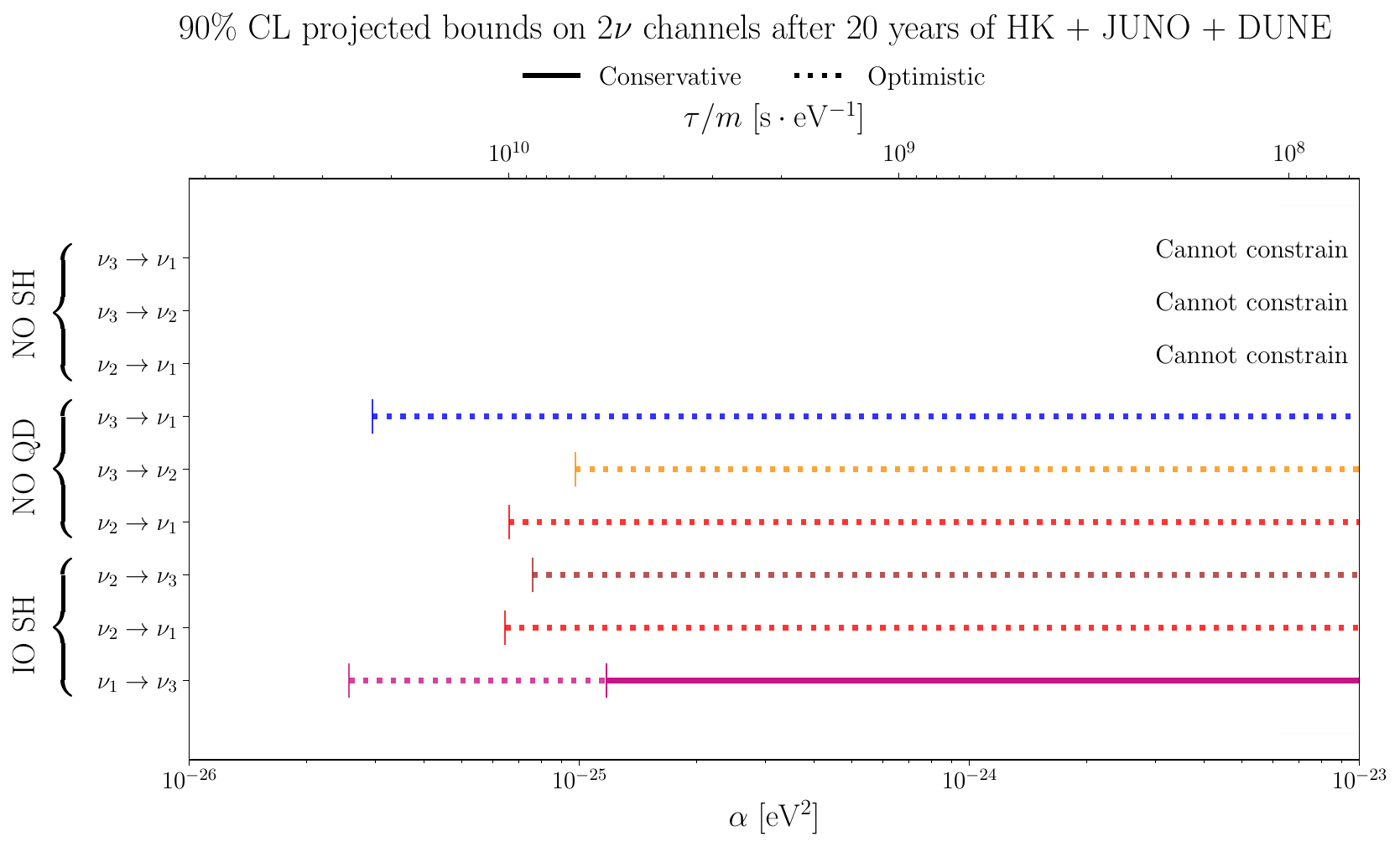}
\caption{\label{fig:2nusensitivities_exclusion_panels} $90\%$ CL projected bounds for all considered $2\nu$ decay channels, assuming 20 years of exposure for Hyper-Kamiokande (HK), JUNO, and DUNE. Bounds in solid lines represent the constraining power under a conservative detection scenario, and bounds in dashed lines represent the constraining power under an optimistic detection scenario (see main text for further details). We do not plot the exclusion bounds for the three  NO SH decay channels because, even under an optimistic detection scenario,  these channels could not be constrained at a $90\%$ CL even after $20$~years. The projected sensitivity to visible neutrino decays would greatly benefit from a significant reduction of the astrophysical uncertainties. 
}
\end{figure}

In Fig.~\ref{fig:2nusensitivities_exclusion_panels}, we present the projected bounds at a $90\%$ CL of Hyper-Kamiokande, JUNO, and DUNE for all considered $2\nu$ decay channels after $20$~years of data taking. We compute the sensitivity of a combined analysis to a decay signature with a $\chi^2$ test, as detailed in Appendix~\ref{sec:statanalysis}. Under a conservative detection scenario, our constraining power is severely limited; we can only constrain the IO SH $\nu_1 \to \nu_3$ channel at a $90\%$ CL when we account for current astrophysical and background uncertainties. Our projected bounds for this channel after $20$~years are
\begin{equation}
    \alpha_1 < 1.18 \times 10^{-25}~\mathrm{eV^2} \qquad \text{or equivalently} \qquad \tau_1/m_1 > 5.61 \times 10^9~\mathrm{s\cdot eV^{-1}}\, .
\end{equation} 

In the optimistic  detection scenario, the constraining power is greatly improved for some channels. If the mass configuration is either NO QD or IO SH, we could exclude complete decay and regions of partial decay for all $2\nu$ channels at a $90\%$ CL. The most stringently constrained channels would be $\nu_3 \to \nu_1$ ($\nu_1 \to \nu_3$) for NO QD (IO SH). However, even with these significant advances in our understanding of  the astrophysics and detector physics, if the mass configuration were NO SH, we could still not constrain any of these decay channels at a $90\%$ CL. 

In Table~\ref{exposuretimestable}, we calculate the exposure time required for a combined  analysis, including Hyper-Kamiokande, JUNO and DUNE, to exclude a given \textit{complete} decay signature, i.e. $\alpha > 10^{-24}$ eV$^{2}$. For NO QD and IO mass configurations, constraints at a $90\%$ C.L. and even a $99\%$ CL are possible in the coming decades with improved understanding of backgrounds and astrophysical uncertainties. In particular, the NO QD $\nu_3 \to \nu_1$ (IO $\nu_1 \to \nu_3$) channels, which increase (decrease) the DSNB flux by a significant overall normalization factor, could be constrained at a $99\%$ CL in less than a decade under our optimistic detection scenario. The exposure time  required to exclude each decay channel at a $90\%$ and $99\%$ CL is  somewhere between the conservative and optimistic cases presented here. It is clear that the effectiveness of the DSNB as a probe for $2\nu$ visible neutrino decay is thus strongly dependent on both the yet unknown neutrino mass ordering, the absolute mass scale, and our  ability to constrain astrophysical models as well as improve our understanding of detectors and backgrounds.

\begin{table}[tbp]
\centering
\begin{tabular}{|c|c|cc|cc|}
\hline
\multicolumn{2}{|c|}{\centering BSM scenario} & \multicolumn{4}{c|}{\centering Exposure Time [years]} \\ \hline
\centering \makecell{Mass\\Configuration} & \centering \makecell{Decay\\Channel} & \makecell{90\% CL\\(conservative)} & \makecell{90\% CL\\(optimistic)} & \makecell{99\% CL\\(conservative)} & \makecell{99\% CL\\(optimistic)} \\ \hline
\multirow{3}{*}{NO SH} & \textcolor{gray}{$\nu_3 \to \nu_1$}& \textcolor{gray}{$>50$} & \textcolor{gray}{$>50$} & \textcolor{gray}{$>50$} & \textcolor{gray}{$>50$} \\  
 & \textcolor{gray}{$\nu_3 \to \nu_2$} & \textcolor{gray}{$>50$} & \textcolor{gray}{$>50$} & \textcolor{gray}{$>50$} & \textcolor{gray}{$>50$} \\ 
 & $\nu_2 \to \nu_1$ & \textcolor{gray}{$>50$} & 22 & \textcolor{gray}{$>50$} & \textcolor{gray}{$>50$} \\ \hline
\multirow{3}{*}{NO QD} & $\nu_3 \to \nu_1$ & \textcolor{gray}{$>50$} & 2 & \textcolor{gray}{$>50$} & 4 \\  
 & $\nu_3 \to \nu_2$ & \textcolor{gray}{$>50$} & 8 & \textcolor{gray}{$>50$} & 22 \\ 
 & $\nu_2 \to \nu_1$ & \textcolor{gray}{$>50$} & 5 & \textcolor{gray}{$>50$} & 13 \\ \hline
\multirow{3}{*}{IO SH} & $\nu_2 \to \nu_3$ & \textcolor{gray}{$>50$} & 5 & \textcolor{gray}{$>50$} & 14 \\  
 & $\nu_2 \to \nu_1$ & \textcolor{gray}{$>50$} & 5 & \textcolor{gray}{$>50$} & 13 \\ 
 & $\nu_1 \to \nu_3$ & 4 & 1 & 15 & 3 \\ \hline\hline
IO & quasi-Dirac & 8 & 2 & \textcolor{gray}{$>50$} & 4 \\ \hline
\end{tabular}
\caption{\label{exposuretimestable} Exposure times required to achieve $90\%$ and 99\% C.L. exclusion to $\alpha = 10^{-24}~\mathrm{eV^2}$ decay assuming a null signal under conservative and optimistic detection scenarios for a combined  analysis involving Hyper-Kamiokande, JUNO and DUNE (see the main text for  details). In the vast majority of scenarios, under conservative assumptions, unrealistic data-taking times would be needed to reach a significant exclusion of the full decay hypothesis. For comparison, we also include the exposure times needed to exclude mass-splittings of quasi-Dirac neutrinos for $\delta m^ 2 = 10^{-21}$~eV$^2$ (see Sec.~\ref{sec:bsm} for a  discussion) for the same conservative and optimistic scenarios.}
\end{table}

As a final remark, previous work  predicts $3\sigma$ exclusion of the NO SH $\nu_3 \to \nu_1$ channel within $20$~years using only Hyper-Kamiokande~\cite{DeGouvea:2020ang}. However, Ref.~\cite{DeGouvea:2020ang} relies on a DSNB model with a  larger integrated flux and a higher energy peak, which both increase the statistics and  the effect that a low-energy pileup would have on the event rate. The authors of Ref.~\cite{DeGouvea:2020ang} also consider fewer background sources and ignore the impact of the fraction of black hole forming collapses in the DSNB. As a result, their projected sensitivities are around an order of magnitude larger than our optimistic forecasts.

\section{Degeneracies between astrophysical uncertainties and beyond the Standard Model physics}
\label{sec:degeneracies}
Different neutrino decay scenarios and other BSM models could lead to comparable signatures in the observed DSNB spectrum. In this section, we investigate the degeneracies between the decay signatures and the astrophysical uncertainties. We then explore the signatures left in the DSNB signal from different decay channels and also focus on exploring the chances of distinguishing visible from invisible neutrino decay, as well as visible decay from quasi-Dirac neutrinos. 

\subsection{Degeneracies between decay signatures and astrophysical uncertainties}

We now discuss how different $2\nu$ decay channels might be degenerate with astrophysical uncertainties, focusing on the supernova rate uncertainty and the $f_\mathrm{BH}$ uncertainty. Because the supernova rate uncertainty is a normalization uncertainty on the DSNB flux, decay channels that predominantly affect the DSNB normalization, i.e.~QD channels and SH channels decaying into $\nu_3$, could result in signatures that are degenerate with the supernova rate. For example, in NO QD, all three channels result in an increase of the $\nu_e$ DSNB  flux normalization that falls within the supernova rate uncertainty band (see Fig.~\ref{fig:DSNBfluxepartialcompletedecay}). We thus expect that our ability to probe $2\nu$ decay channels is enhanced the most by improvements in our knowledge on the supernova rate normalization.

The uncertainty in $f_\mathrm{BH}$ primarily affects the spectral shape of the DSNB. We thus expect that spectrum-distorting decay channels, such as $\nu_2 \to \nu_1$ (NO SH) and  partial decay channels, are  most degenerate with $f_\mathrm{BH}$. Due to the large effect the supernova rate uncertainty has on the DSNB, we expect that the $f_\mathrm{BH}$ uncertainty should have a subdominant impact on our sensitivity to even these channels. However, if significant progress is made on pinpointing the supernova rate normalization, the $f_\mathrm{BH}$ uncertainties should  become more relevant when attempting to probe spectrally distorting neutrino decays. While the supernova rate can be measured through multi-wavelength electromagnetic data~\cite{Madau:2014bja,Ekanger:2023qzw}, probing the black hole fraction could take great advantage from the DSNB detection~\cite{Lien:2010yb,Moller:2018kpn,Kresse:2020nto}, in addition to  searches for disappearing red supergiants~\cite{Neustadt:2021jjt}.

\subsection{Degeneracies between different decay channels}

Several decay scenarios result in very similar qualitative modifications on the electron-flavor component of the DSNB, especially for energies above $10$~MeV. For instance, all NO SH decay channels distort the spectrum, and all remain quite degenerate with the null signal in the first place. NO QD and IO SH decay channels primarily affect the DSNB normalization, and some of the channels could thus be degenerate with each other.

To quantify the degree to which we could distinguish different visible decay channels using the DSNB, we employ the following process. For two decay channels, A and B, we compute two $\chi^2$ values using the statistical analysis outlined in Appendix~\ref{sec:statanalysis}; one assuming scenario $A$ is true, $\Delta\chi^2_{A\mathrm{~true}}(B)$, and one assuming scenario $B$ is true, $\Delta\chi^2_{B\mathrm{~true}}(A)$. We then define
\begin{equation}
    \label{eqn:discriminatingdecays}
    \Delta\chi^2_\mathrm{disc}(A, B) = \mathrm{min}\lbrace\Delta\chi^2_{A\mathrm{~true}}(B), \Delta\chi^2_{B\mathrm{~true}}(A)\rbrace
\end{equation}
as our ability to statistically discriminate the two decay channels. 

\begin{figure}[tbp]
\centering 
\includegraphics[width=\linewidth]{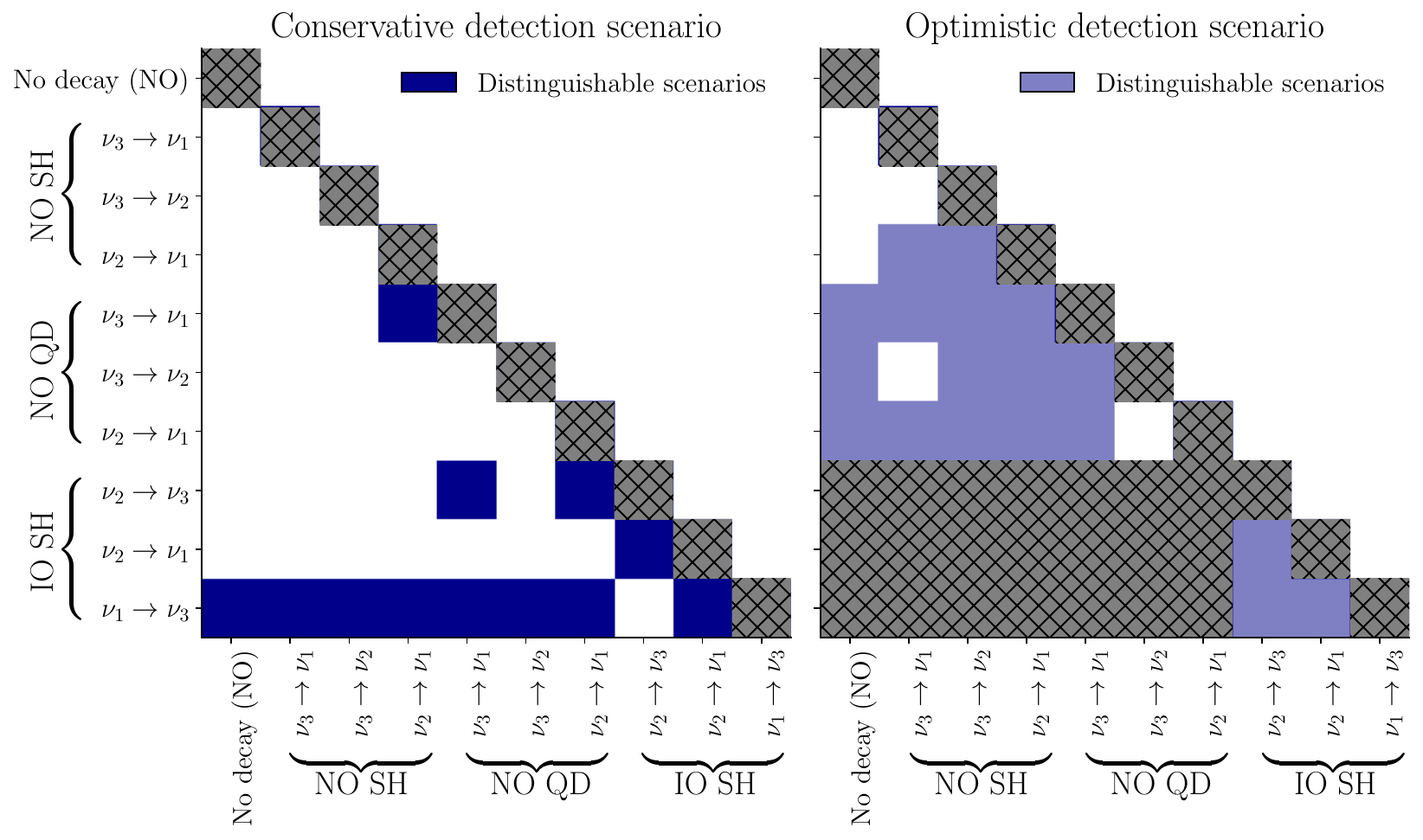}
\caption{\label{fig:2nu_confusion_plot} Pairs of neutrino decay channels whose signatures could be discriminated at a $90\%$ CL using $20$ years of data from Hyper-Kamiokande, JUNO and DUNE in a combined analysis. Pairs in dark blue can be discriminated after $20$ years under our conservative detection scenario; pairs in light blue can be discriminated after $20$ years under our optimistic detection scenario. The diagonals of both panels are hatched out because they assess how well we could discriminate between identical decay channels. The bottom left rectangle in the right panel is hatched out because it compares NO and IO decay channels against each other; we expect to have knowledge of the mass ordering after $20$~years and thus consider comparing these pairs irrelevant under our optimistic detection scenario.}
\end{figure}

In Fig. \ref{fig:2nu_confusion_plot}, we identify the decay channels we can discriminate at a $90\%$ CL after $20$~years 
(i.e.~channels $A$ and $B$ for which $\Delta\chi^2_\mathrm{disc}(A, B) \geq 2.71$). As evidenced in the left panel, under a conservative detection scenario assuming current knowledge of astrophysics and detector physics, the vast majority of the DSNB decay signatures are degenerate among themselves. Even in the $\nu_1 \to \nu_3$ IO SH channel, which is the only channel we can constrain at a $90\%$ CL under a conservative detection scenario, we find that this signature would be degenerate with the $\nu_2 \to \nu_3$ IO SH channel. Hence, under conservative assumptions, we would not be able to pinpoint the decay channel responsible for a distortion in DSNB spectrum with statistical significance.

This picture would improve significantly if the astrophysical uncertainties on the supernova rate  and the fraction of black hole forming collapses were reduced to negligible levels and if neutral-current atmospheric backgrounds at Hyper-Kamiokande could be rejected. Additionally, provided that the mass ordering and the absolute mass scale were determined by independent means, all decay channels in IO  could be distinguished, as well as $\nu_2 \to \nu_1$ for NO SH masses and $\nu_3\to\nu_1$ for NO QD masses (see right panel in Fig.~\ref{fig:2nu_confusion_plot}). 

We outline the  degeneracies with respect to the three-neutrino ($3\nu$) decay framework  in Appendix~\ref{sec:DSNB3nudecay}. The increased number of free parameters and varied phenomenology result in many degeneracies between decay scenarios; however, there are a few key regimes that are useful to point out. In the regime where both of the heavier mass eigenstates completely decay, most $3\nu$ scenarios become almost completely degenerate with each other, as at Earth all neutrinos will be in the lightest mass eigenstate regardless of the branching ratios. In addition, assuming that the lifetime of one of the heavier mass eigenstates approaches zero, the decay signature of the $3\nu$ scenario becomes degenerate with a $2\nu$ scenario, as effectively only one of the heavier mass eigenstates decays.

\subsection{Degeneracies between visible neutrino decay and other beyond the Standard Model scenarios}
\label{sec:bsm}

Under an optimistic detection scenario, our ability to discriminate between decay channels can grow significantly. However, these still remain degenerate with other BSM  scenarios. In this section, we investigate whether it is possible to discriminate among different BSM scenarios, focusing on invisible neutrino decay and quasi-Dirac neutrinos. 

\begin{figure}[tbp]
\centering 
\includegraphics[width=\linewidth]{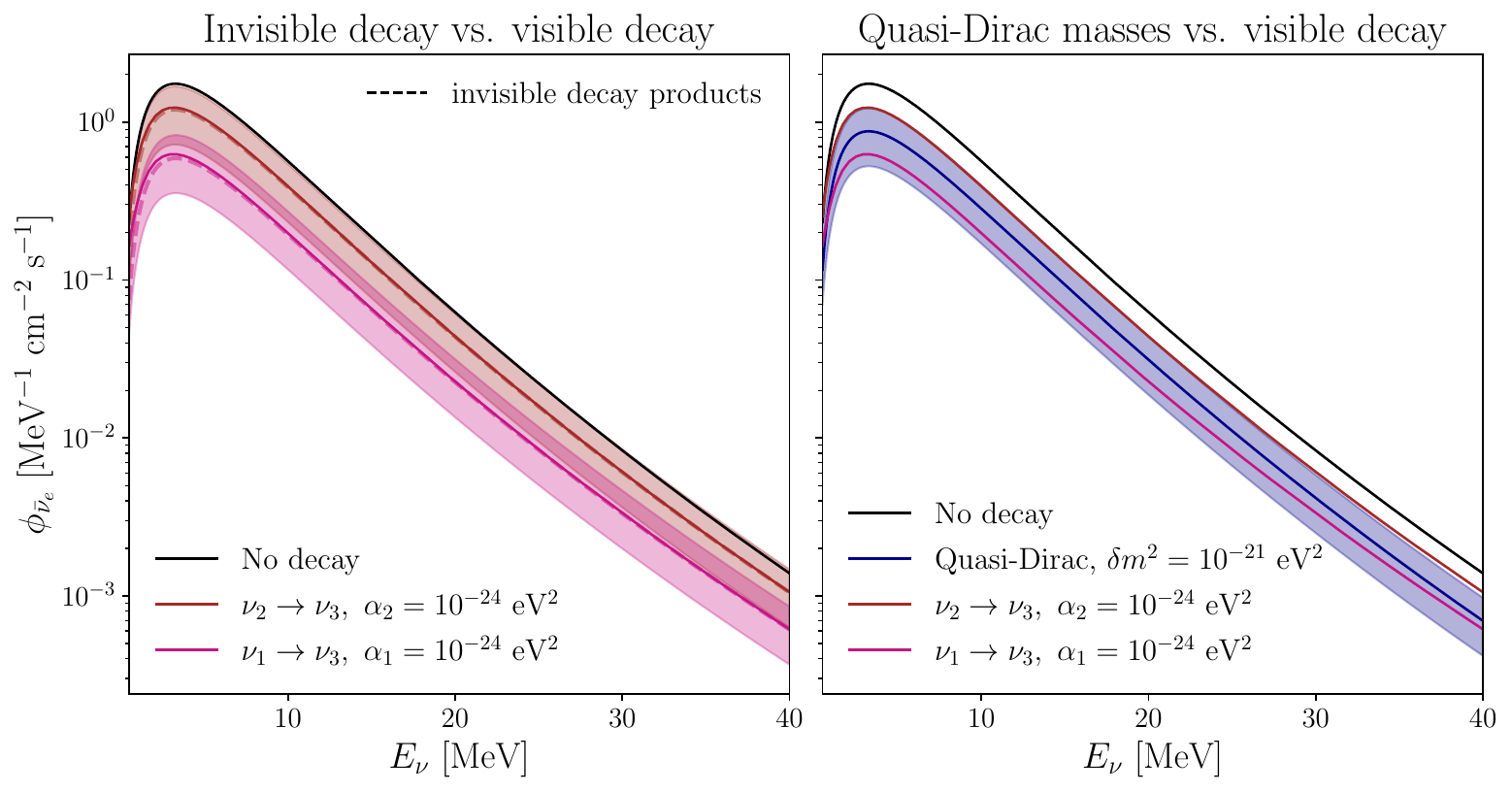}
\caption{\label{dsnbdegensfig} Electron antineutrino DSNB flux expected at Earth (for IO  and $m_{\nu_3}\sim 0$ eV) as a function of the observed neutrino energy for different BSM scenarios. The  DSNB forecasted in the absence of BSM physics is plotted in black. \textit{Left panel:} Visible (solid) and invisible (dashed) $\nu_1$ and $\nu_2$ decays are plotted in magenta and brown, respectively.
\textit{Right pannel:}  Comparison between visible neutrino decays ($\nu_2 \to \nu_3$ and $\nu_1 \to \nu_3$, brown and magenta lines) and averaged-out quasi-Dirac neutrinos oscillations (blue). Invisible neutrino decays and averaged-out quasi-Dirac neutrino oscillations are degenerate with some visible neutrino decay scenarios.}
\end{figure}

\subsubsection*{Invisible neutrino decay}

We have focused on the possibility that neutrino decay products are, at least in part, visible; however, another possibility is that the neutrino decay products are not detectable, either because they are feebly interacting or have too little energy. The dominant signature of invisible decay at the DSNB is a decrease in the flux, starting at low energies for partial decays and resulting in a normalization decrease for complete decay~\cite{Martinez-Mirave:2024hfd}. 
A simple scenario of invisible neutrino decay would result from Dirac neutrinos coupling to a lwpton-number zero scalar~\cite{deGouvea:2019goq}. Depending on the structure of the couplings and whether parity is conserve, the decay products can be right-handed neutrinos (left-handed antineutrinos) and therefore, effectively invisible.
If the mass ordering was IO, due to the smallness of $|U_{e3}|^2$, decay from a heavier mass state into $\nu_3$ results in a very similar signature to decay from the given heavier mass state into an invisible product. The signals of $\nu_2 \to \nu_3$ or $\nu_1 \to \nu_3$ and $\nu_2 \to \mathtt{invisible}$ or $\nu_1 \to \mathtt{invisible}$ respectively would thus be almost completely degenerate, and effectively indistinguishable from each other, even with perfect knowledge of the DSNB flux and our detector physics. We show two cases of $2\nu$ visible vs. invisible decay in Fig. \ref{dsnbdegensfig} to illustrate this fact. 

Note that for certain visible and invisible decay signatures, such as $\nu_2 \to \nu_3$ and $\nu_2 \to \mathtt{invisible}$ signatures, even with near-perfect knowledge of astrophysical uncertainties it will be near-impossible to break the visible-invisible decay degeneracies with only the DSNB $\nu_e$ and $\bar{\nu}_e$ fluxes. A flavor-insensitive measurement of the DSNB would be needed to shed light on the matter, a scenario which is probably not possible in the near future~\cite{Suliga:2021hek}.

In Fig. \ref{fig:BSM_confusion_plot}, we compare how well we could discriminate between invisible decay channels and IO visible decay channels under an optimistic scenario. Even with negligible astrophysical uncertainties and greatly improved background reduction, $\nu_i \to \nu_3$ and $\nu_i \to \mathtt{invisible}$ channels remain degenerate. These results are particularly notable because the IO mass configuration represents our best chance at constraining visible decay, as shown in Fig. \ref{fig:2nusensitivities_exclusion_panels}. While it may be possible to set cutting-edge bounds on these visible decay channels assuming a detection consistent with no decay, in the event that the DSNB detection yields results consistent with a visible decay signature, it can be challenging to infer the true nature of the BSM physics involved.

\subsubsection*{Quasi Dirac neutrinos}
The Dirac or Majorana nature of neutrinos depends on the underlying mechanism responsible for neutrino masses. The so-called quasi-Dirac neutrino scenario arises when the Lagrangian includes both Majorana and Dirac mass terms, but the Majorana terms are small, yet non-zero~\cite{Valle:1982yw,Chang:1999pb,Sanchez:2001ua,Stephenson:2004wv,McDonald:2004qx,Abel:2004tt,Ahn:2016hhq,Fonseca:2016xsy}. Although small, these terms can result in the existence of new mass splittings, mixing angles, and phases in the lepton mixing matrix~\cite{Anamiati:2017rxw}.

We do not aim to explore the rich phenomenology resulting from the large number of free parameters in this scenario. Instead, we consider that the mixing is unaltered and have three active-sterile mass state pairs, each pair having almost degenerate masses separated by three $\delta m_i^2$. These result in ultra-long-baseline scale oscillations. The probability that a mass state $\nu_i$ can oscillate into a flavor state $\nu_\alpha$ at Earth from some redshift $z$ and energy $E_\nu$ is~\cite{DeGouvea:2020ang}:
\begin{equation}
    \label{eq:quasiDirac_pheno}
    P_{i\alpha}(z, E_\nu) = \frac{1}{2}|U_{\alpha j}|^2\left[] 1 + e^{-\mathcal{L}_j^2(z)}\cos\left(\frac{\delta m_j^2 \Upsilon(z)}{2E_\nu}\right)\right]\, ,
\end{equation}
with $\Upsilon(z)$ defined as in Eq.~\eqref{eq:effectivelength} and where the decoherence factor,
\begin{equation}
    \label{Ldecoh}
    \mathcal{L}_j(z) = \frac{\delta m_j^2}{4\sqrt{2} E_\nu^2 \sigma_x}\int_0 ^z \frac{\text{d}z'}{(1+z')^3 H(z')}\, ,
\end{equation}
accounts for the loss of coherence over cosmological distances and it depends on the initial size of the wavepacket ($\sigma_x$).

For neutrino oscillations to be visible, sources would need to be at a redshift such that~\cite{Beacom:2003nk,Esmaili:2012ac}
\begin{align}
    \Upsilon(z) \sim \frac{4\pi}{\delta m^2_j}\approx 8.03~\mathrm{Gpc}~\left(\frac{E_\nu}{10~\mathrm{MeV}}\right)\left(\frac{10^{-25}~\mathrm{eV^2}}{\delta m_j^2}\right)\, . 
\end{align}
For larger redshifts, neutrino oscillations would be averaged out. Moreover, the oscillations are also damped for large redshift due to decoherence. 

Current upper limits on quasi-Dirac mass splitting ($\delta m^2 < 10^{-9}~\mathrm{eV^2}$) are placed by solar neutrinos~\cite{deGouvea:2009fp, Ansarifard:2022kvy}, with sparse regions of the parameter space being constrained down to $\delta m^2 \sim 10^{-18}~\mathrm{eV^2}$~\cite{Rink:2022nvw} using the detection of neutrinos from the active galaxy NGC 1068~\cite{IceCube:2022der} and $\delta m^2 \sim 10^{-20}~\mathrm{eV^2}$~\cite{Martinez-Soler:2021unz} using neutrino data from the supernova SN 1987A. Recent studies~\cite{Carloni:2022cqz, Fong:2024mqz} have demonstrated the possibility of using other astrophysical neutrino sources to constrain parameter space down to $\delta m^2 \sim 10^{-20}~\mathrm{eV^2}$ with IceCube and IceCube-Gen2~\cite{IceCube-Gen2:2020qha}. 

In principle, the detection of the DSNB could further improve these prospects since neutrinos travel over cosmological distances and have smaller energies, i.e.~$\mathcal{O}(10)$~MeV. In particular, for the DSNB, for mass-splittings of $\delta m^2 > 10^{-22}$~eV$^2$ and initial wavepacket sizes $\sigma_x \sim 10^{-10}$--$10^{-15}$~m, flavor oscillations are damped and averaged out; then the probability in Eq.~\eqref{eq:quasiDirac_pheno} becomes
\begin{align}
     P_{i\alpha}(z, E_\nu) = \frac{1}{2}|U_{\alpha j}|^2\,.
\end{align}
As a consequence, the observable signature in this regime is a $50\%$ reduction of the DSNB flux. 

Considering all active-sterile mass splittings to be equal, for this range of the parameter space ($\delta m^2 > 10^{-22}$ eV$^2$), we see that quasi-Dirac oscillations can result in a signature which is qualitatively very similar to IO decays to $\nu_3$. We present IO visible decay cases along with an averaged out quasi-Dirac case in Fig.~\ref{dsnbdegensfig}. The supernova rate  normalization uncertainty band contains complete decay signatures of both $\nu_2 \to \nu_3$ and $\nu_1 \to \nu_3$ visible decay. The quasi-Dirac and decay signatures are not completely degenerate, and the degeneracy is mainly a product of the extant astrophysical uncertainties. Increasing precision measurements of the supernova rate would in this specific case be invaluable in breaking the degeneracies between these two different BSM models.

\begin{figure}[tbp]
\centering 
\includegraphics[width=\linewidth]{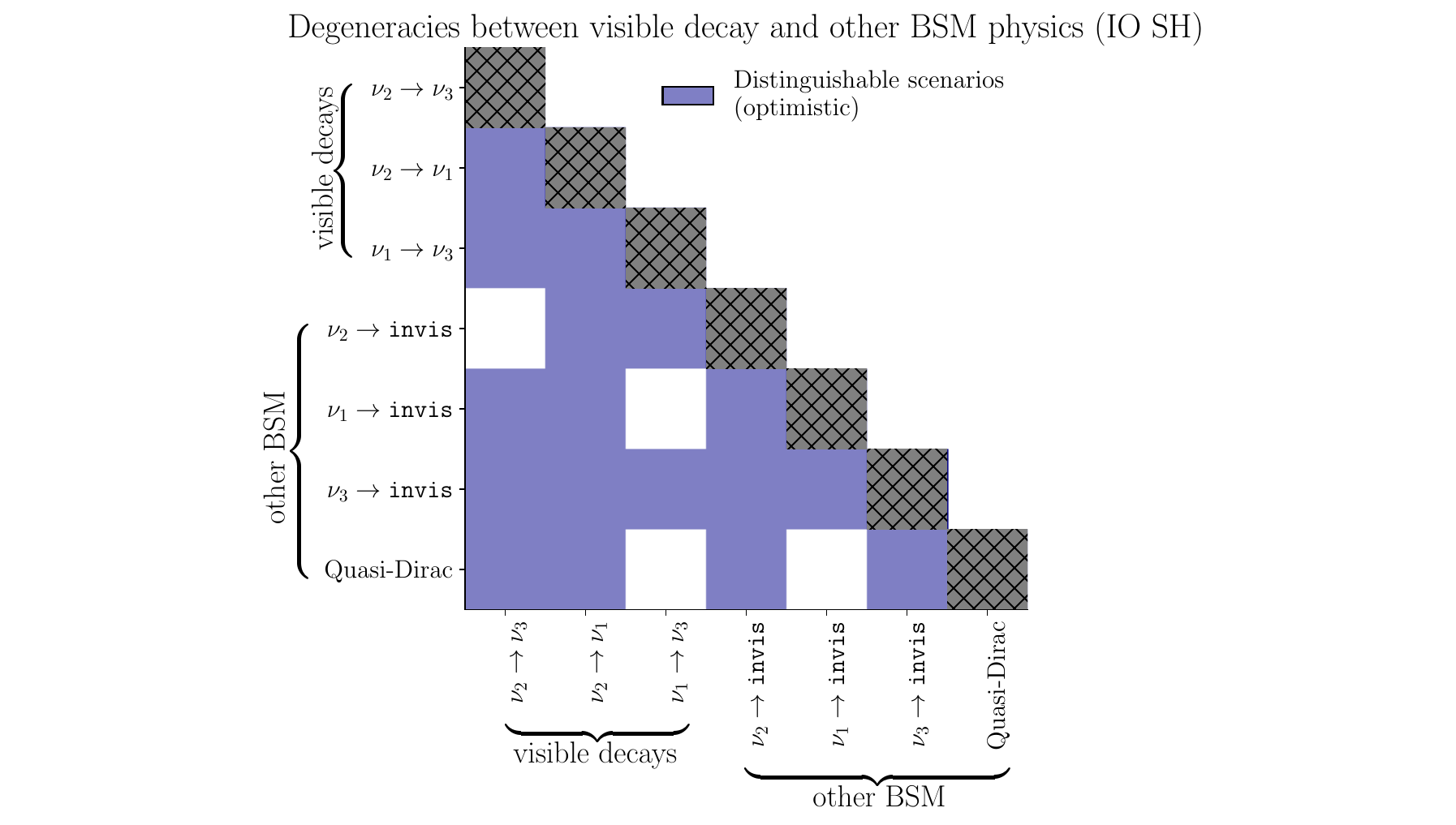}
\caption{\label{fig:BSM_confusion_plot} Pairs of BSM scenarios (IO SH visible decays, invisible decays, and averaged-out quasi-Dirac neutrino oscillations) whose signatures could be discriminated at a $90\%$ CL using $20$ years of data from Hyper-Kamiokande, JUNO and DUNE in a combined analysis. The diagonal is hatched out because these boxes assess how well we could discriminate between identical decay channels.}
\end{figure}

In Fig.~\ref{fig:BSM_confusion_plot}, we display how well we could discriminate between an averaged-out quasi-Dirac signature and various visible and invisible decay signatures in an IO SH mass configuration. Like with invisible decays, we see that even under an optimistic detection scenario, the quasi-Dirac signature in the DSNB is still very degenerate with other BSM scenarios; while an improved supernova rate normalization knowledge breaks the degeneracy between quasi-Dirac and $\nu_2 \to \nu_3$ shown in Fig.~\ref{dsnbdegensfig}, we could not discriminate it at a $90\%$ CL from $\nu_1$ decays to either $\nu_3$ or invisible products.

It is important to point out that while we have explored in detail the level of degeneracy in the DSNB between a few BSM scenarios, additional  BSM physics  could modify neutrino propagation, leaving an observable imprint on the DSNB (e.g, spin-flavor conversion of Dirac neutrinos due to nonzero magnetic moments would also manifest as a reduction of the observable DSNB flux~\cite{Kopp:2022cug}).

\section{Outlook}
\label{sec:conclusion}

The diffuse supernova neutrino background (DSNB) accounts for all neutrinos emitted by core-collapse supernovae throughout cosmic history. The DSNB  holds a wealth of information about both the supernova population and the neutrinos themselves. The detection of the DSNB is on the  near horizon, with hints already showing up in  the data of the Gadolinium-doped Super-Kamiokande water Cherenkov detector~\cite{SK-Gd:2024}. Several extant astrophysical uncertainties constrain  our current knowledge of the DSNB flux, leading to normalization uncertainties  and spectral uncertainties in the slope of the high energy tail. Gaining precision information about the DSNB flux is further complicated by a plethora of backgrounds at neutrino observatories in the same energy range as the DSNB. In the coming decades, next generation neutrino telescopes will not only detect the DSNB at high significance, but will also collect enough statistics for us to glean invaluable information about the characteristics of the DSNB flux. 

In this work, we examine the possibility of probing visible neutrino decay using the detection of the DSNB in the next-generation neutrino observatories Hyper-Kamiokande, JUNO, and DUNE. We forecast our sensitivity to the parameter space of visible neutrino decay, keeping our assumptions as model agnostic as possible and considering single channels and combined multi-channel decay scenarios. In our statistical analysis, we incorporate astrophysical uncertainties in the core-collapse supernova rate and the fraction of black hole forming collapses, as well as the experimental backgrounds plaguing the DSNB event rate. We present sensitivity forecasts with two sets of assumptions: a conservative case, where we assume current knowledge about the supernova rate, the fraction of black hole forming collapses, and backgrounds, and an optimistic case, where we assume $5\%$-level uncertainties of the astrophysics and backgrounds and excellent background-removal capabilities.

We find that the DSNB phenomenology  due to neutrino decay is strongly dependent on the still-unknown neutrino mass ordering and absolute mass scale. If the neutrino mass ordering is normal (NO) and the lightest mass state is approximately zero, neutrino decay marginally affects the DSNB flux, even in scenarios where all heavy mass states decay. If the mass ordering is normal and the absolute mass scale is large, the predominant decay signature is an upward shift in the DSNB normalization. If the neutrino mass ordering is inverted (IO), the phenomenology is very rich and diverse, although in complete decay we see a drop in the DSNB normalization.

While visible neutrino decay manifests a rich phenomenology in the DSNB flux, our ability to constrain decay channel lifetimes is severely hampered by our incomplete knowledge of the population of collapsing massive stars.
Some of the main  sources of uncertainty in the DSNB flux (the  normalization of the supernova rate and the fraction of black hole forming collapses) are degenerate with normalization-altering and spectrum-distorting features due to neutrino decay, respectively. Under a conservative detection scenario, we find that we can only constrain specific  decay scenarios, assuming IO, to any appreciable statistical significance. Under our optimistic detection scenario, where we effectively remove all astrophysical uncertainties, we are able to probe most IO decay channels and NO decay channels if the absolute mass scale is large, i.e. $m_\text{lightest}\sim 0.35$--$0.45$~eV. The decay signatures under NO with a near-zero absolute mass scale remain too degenerate with the no-decay DSNB signal to learn anything about neutrino lifetimes.

We further demonstrate that the DSNB flux under visible neutrino decay is not only degenerate with poorly constrained astrophysics; it is also degenerate with other Beyond-Standard-Model (BSM) physics scenarios. In some cases, like quasi-Dirac neutrino oscillations, these degeneracies can be broken by improving our astrophysical knowledge. In other scenarios, like neutrino invisible decays, the signatures are so similar that even perfect knowledge of the astrophysics would not help in discriminating these signals. This simple exercise quantatively points to the challenges  of constraining or discovering BSM physics relying on the DSNB. Even if we were to have perfect knowledge of all our uncertainties and discovered a DSNB flux consistent with one BSM model, it is  possible that another  different BSM model results in a  degenerate signature, and it is hard to gain conclusive information about the actual nature of the neutrino physics behind the scenes.

Our findings illustrate that, while the DSNB can  be a powerful tool for probing BSM  physics, its effectiveness  over the coming decades is strongly dependent on 
an improvement of the experimental techniques to eliminate the backgrounds at next generation neutrino observatories as well as advances in our understanding of the astrophysical uncertainties entering the modeling of the DSNB.  
With building evidence that  the  DSNB detection is becoming a reality, it is now more important than ever to  realistically quantify what information we can gather from such a unique source.

\acknowledgments

This project has received support from the Villum Foundation 
(Project No.~13164), the Danmarks Frie Forskningsfond (Project No.~8049-00038B), the  Carlsberg Foundation (CF18-0183) and the Deutsche Forschungsgemeinschaft through Sonderforschungbereich SFB 1258 ``Neutrinos and Dark Matter in Astro- and Particle Physics'' (NDM). MM acknowledges support from the Herchel Smith Undergraduate Science Research Program. The Tycho supercomputer hosted at the SCIENCE HPC Center at the University of Copenhagen was used for supporting the numerical simulations presented in this work.

\appendix
\section{Event rates in next generation neutrino observatories}

\label{sec:DSNBdetection}

The detection of the DSNB is among the milestones of next-generation neutrino observatories. In this work, we focus on the sensitivity of Hyper-Kamiokande~\cite{Hyper-Kamiokande:2016srs,Hyper-Kamiokande:2018ofw},  JUNO~\cite{JUNO:2015zny,JUNO:2021vlw}, and DUNE~\cite{DUNE:2020ypp} to the electron-flavor component of the DSNB. 
 
In this appendix, we describe the modeling of the event rates at Hyper-Kamiokande, JUNO, and DUNE. We also discuss the background sources at each detector. 
 
\subsubsection*{Hyper-Kamiokande}
The water Cerenkov neutrino experiment Hyper-Kamiokande \cite{Hyper-Kamiokande:2018ofw} will detect the electron antineutrinos from the DSNB through inverse beta decay (IBD,  $\bar{\nu}_e + p \to e^+ + n$). We  model the expected event rate as
\begin{equation}
    \label{eq:HK_event_rate}
    R_\mathrm{HK} = \tau N_t \varepsilon_\mathrm{IBD} \int_0^\infty dE'_e~\sigma_\mathrm{IBD}(E_\nu, E'_e)\Phi_{\bar{\nu}_e}(E_\nu)\int_{E_{e,\mathrm{min}}}^{E_{e,\mathrm{max}}} dE_e~K(E'_e, E_e)\, ,
\end{equation}
where $\tau$ denotes the exposure time, $N_t$ is the number of targets, and $\varepsilon_\text{IBD}$ is the detection efficiency. Here, $E_\nu$, $E'_e$, and $E_e$ are the true neutrino energy, the true positron energy, and the reconstructed positron energy, respectively. We consider two tanks of $187$~kt of water loaded with Gadolinum, so that $N_t = 2.5 \times 10^{34}$ and $\varepsilon_\text{IBD} = 0.67$~\cite{Hyper-Kamiokande:2018ofw}. In Eq.~\eqref{eq:HK_event_rate}, $\sigma_\mathrm{IBD}$ is the IBD cross section~\cite{Strumia:2003zx,Ricciardi:2022pru} and the innermost integral takes into account the energy resolution. We consider a Gaussian response function: 
\begin{equation}
    \label{eq:IBD_response_function}
    K(u, v) = \frac{1}{\delta(v)\sqrt{2\pi}}e^{-\left(\frac{u-v}{\delta(v)\sqrt{2}}\right)^2}\, ,
\end{equation}
with
\begin{equation}
\label{eq:IBD_eres}
    \delta(E_e) = 0.1\sqrt{E_e [\mathrm{MeV}]}\, .
\end{equation}
Finally, $E_{e,\mathrm{min}}$ and $E_{e,\mathrm{max}}$ are the minimum and maximum reconstructed positron energy for each energy bin.

Knowledge of the detector backgrounds is essential to  shape the energy window for DSNB searches. For Hyper-Kamiokande, reactor antineutrinos dominate the flux below  $\sim 10~\mathrm{MeV}$. The dominant background at energies between $10$ and $20$~MeV is the neutral current  background caused by atmospheric neutrino quasi elastic scattering~\cite{Kunxian:2016joi,Ashida:2020erk}, although  this background could be reduced significantly in the future~\cite{Maksimovic:2021dmz}. Hence, we report our results with and without this background. Backgrounds from $^9\mathrm{Li}$ spallation~\cite{Hyper-Kamiokande:2018ofw} are also present for energies between $10$  and $20$~MeV. Additionally, ``invisible'' atmospheric muons and (to a lesser extent) atmospheric CC $\nu_e$ flux start to overwhelm the DSNB signal at energies above $ 20$~MeV~\cite{Hyper-Kamiokande:2018ofw}. Hence, we limit our analysis to reconstructed positron energies between $10$  and $32$~MeV.

Note that while we focus on  Hyper-Kamiokande in this work,  plans for the next-to-next generation neutrino experiment ESSnuSB are currently under development~\cite{Abel:2004tt,ESSnuSB:2023ogw}. Studies for the far detector site featuring two water-cherenkov detectors of $270$~kt each loaded Gadolinium are foreseen. Such experimental efforts would positively impact the DSNB  new physics studies, like the ones  presented here.

\subsubsection*{JUNO}
JUNO~\cite{JUNO:2015zny}, a liquid-scintillation detector with a fiducial volume of $17$~kt, will  be sensitive to electron antineutrinos from the DSNB via IBD. We compute the expected event number as in Eq.~\eqref{eq:HK_event_rate},  for a number of targets of $N_t = 1.2 \times 10^{33}$. We assume a $50\%$ detection efficiency~\cite{Martinez-Mirave:2024hfd}, although it might be possible to reach efficiencies of up to $80\%$~\cite{Cheng:2023zds}. We take the same energy resolution for JUNO as we do for Hyper-Kamiokande (cf.~Eq.~\ref{eq:IBD_eres}). The dominant backgrounds in the DSNB detection window for JUNO are reactor antineutrinos and atmospheric neutral current and charged current backgrounds~\cite{JUNO:2015zny,Mollenberg:2014pwa,Cheng:2023zds}. Hence, for JUNO, we consider an energy window between $10$ and $34$~MeV.

\subsubsection*{DUNE}
DUNE \cite{DUNE:2020ypp}, a $40$~kt liquid Argon neutrino telescope, could also detect the DSNB through the charged current  reaction $\nu_e + ^{40}\mathrm{Ar} \to e^- + ^{40}K^*$.  The expected rate is
\begin{equation}
    R_\mathrm{DUNE}= \tau N_t \varepsilon_\mathrm{DUNE} \int_0^\infty dE_\nu~\sigma_{\nu_e\mathrm{CC}}(E_\nu) \Phi_{\nu_e}(E_\nu) \int_{E_{r,\mathrm{min}}}^{E_{r,\mathrm{max}}}dE_r~K_\mathrm{DUNE}(E_\nu, E_r)\, ,
\end{equation}
where here $N_t = 6.02 \times 10^{32}$, $\sigma_{\nu_e\text{CC}}$ is the scattering cross section~\cite{snowglobes}, and we consider a detection efficiency $\varepsilon_\text{DUNE} =0.86$~\cite{Moller:2018kpn}.
We parametrize the response function of the detector, which relates the true and reconstructed neutrino energy ($E_\nu$ and $E_r$ respectively), as in Eq.~\eqref{eq:IBD_response_function} with $\delta_\mathrm{DUNE}(E_\nu) = 0.2 E_\nu$~\cite{DUNE:2020ypp,Castiglioni:2020tsu}.

Atmospheric charged current neutrinos constitute a large background at energies above $ 29~\mathrm{MeV}$~\cite{Cocco:2004ac}, whereas for energies below $19~\mathrm{MeV}$, the signal is dominated by solar neutrinos. Based on these two backgrounds, we define our detection window to be between $19$  and $31$~MeV. 

\section{Statistical analysis for effective two-neutrino decays}
\label{sec:statanalysis}

To quantify the statistical significance of our results, we employ a frequentist approach and use a simple $\chi^2$ test. We  define our $\chi^2$ function for each experiment as
\begin{equation}
\label{eq:chi2func}
    \begin{aligned}
        \chi^2(f_\mathrm{BH}, \xi, \eta_1, \dots, \eta_n) &= 2\sum_i \left\{
        (1+\xi)N_i(f_\mathrm{BH}) + \sum_n \eta_nB_{n,i} - S_i\right. \\
        &+ \left(S_i \left.+ \sum_nB_{n,i}\right)\log\left(\frac{S_i + \sum_n B_{n,i}}{(1+\xi)N_i(f_\mathrm{BH}) + \sum_n (1+\eta_n)B_{n,i}}\right)\right\} \\
        &+ \left(\frac{\xi}{\sigma_{R_\mathrm{{SN}}}}\right)^2 + \sum_n \left(\frac{\eta_n}{\sigma_n}\right)^2\, ,
    \end{aligned}
\end{equation}
where $N_i$ represents the expected number of DSNB events under a decay scenario in the $i$th energy bin, $S_i$ are the expected number of DSNB events (we take $f_\mathrm{BH} = 0.21$ to be the ``true'' value) with no decay, and $B_{n, i}$ are the expected background events from the $n$th background source. The fraction of black hole forming collapses, $f_\mathrm{BH}$, is left as a free parameter within the range $[0.09, 0.41]$. We introduce the pull parameter to quantify the role of the supernova rate uncertainty and set $\sigma_{R_\mathrm{SN}} = 0.4$ in accordance with the normalization choice in Sec.~\ref{sec:modelDSNBdecay}. For each of the $n$ background sources, we include an additional pull parameter with $\sigma_n = 0.2$ and minimize the $\chi^2$ function with respect to these parameters. Note that, when reporting the results of combined analysis, the pull parameters accounting for the DSNB uncertainties  ($\xi$ and $f_{\rm BH}$) are common for all the experiments under consideration.

\section{Generalizing two body visible neutrino decay }
\label{sec:DSNB3nudecay}

In Sec.~\ref{sec:DSNB2nudecay}, we outline  how neutrino decay in the effective two-neutrino framework affects the DSNB flux. In this appendix, we explore a  three-neutrino scenario. In this case, the phenomenology of neutrino decays is extremely broad and complex, since several decay channels can be allowed simultaneously with different branching ratios (see Ref.~\cite{Maltoni:2008jr,deGouvea:2022cmo} for several model-dependent examples). Hence, a generic investigation of the allowed parameter space without underlying assumptions is challenging and not promising due to the many degeneracies in the picture. The common approach is to rely on some assumptions, such as two of the neutrinos being stable--i.e.~the effective two neutrino framework--or all branching ratios being equal~\cite{Fogli:2004gy}.

We consider two illustrative limiting $3\nu$ cases, which  demonstrate the broad phenomenology of a $3\nu$ decay scenario and refer to the heaviest, lightest and intermediate mass eigenstates as $\nu_\mathrm{heaviest}$, $\nu_\mathrm{lightest}$, and $\nu_\mathrm{middle}$, respectively. Both cases are built out of  two $2\nu$ channels, which has the advantage of leaving the branching ratios as fixed parameters. 
\begin{enumerate}[i.]
    \item In the first case, we consider the $\nu_\mathrm{heaviest} \to \nu_\mathrm{lightest}$ and $\nu_\mathrm{middle} \to \nu_\mathrm{lightest}$ channels, assuming no other channels are allowed (i.e. their respective branching ratios are zero). 
    \item The second case  assumes only $\nu_\mathrm{heaviest} \to \nu_\mathrm{middle}$ and $\nu_\mathrm{middle} \to \nu_\mathrm{lightest}$ are possible. This case allows mass states to ``pile up'' in the middle mass state, slowing the rate at which neutrinos completely decay into the lightest mass state.
\end{enumerate}

\begin{figure}[tbp]
\centering 
\includegraphics[width=\linewidth]{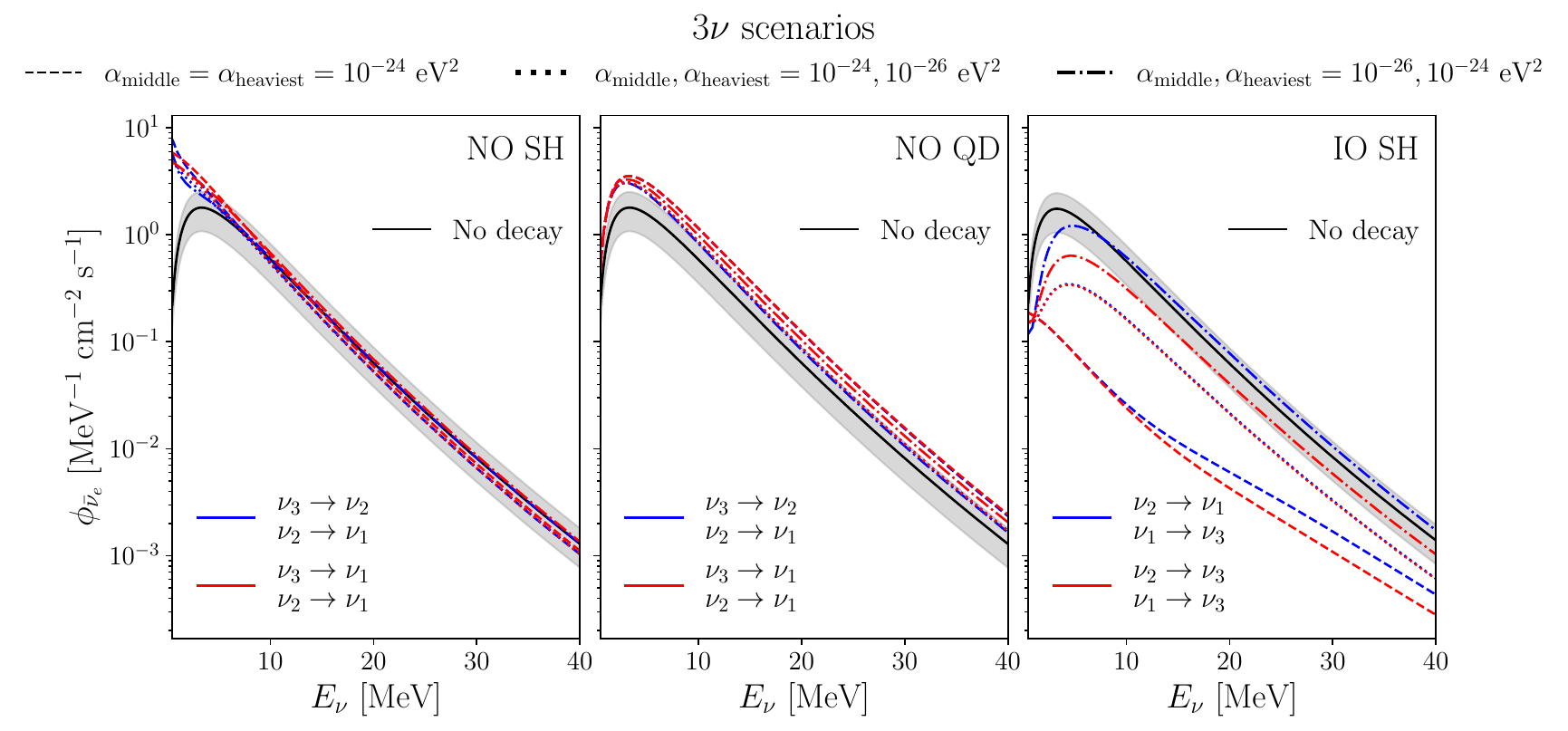}
\caption{\label{3nudecaychannelsfig} Electron antineutrino component of the DSNB flux under two representative $3\nu$ decay scenarios as a function of the antineutrino energy $E_\nu$. The $\nu_\mathrm{heaviest} \to \nu_\mathrm{middle} + \nu_\mathrm{middle} \to \nu_\mathrm{lightest}$ ($\nu_\mathrm{heaviest} \to \nu_\mathrm{lightest} + \nu_\mathrm{middle} \to \nu_\mathrm{lightest}$) decay scenario is plotted in blue (red). We plot three different decay parameter values: one where both heavy mass states completely decay (i.e.~$\alpha_\mathrm{heaviest} = \alpha_\mathrm{middle} = 10^{-24}~\mathrm{eV^2}$, plotted as the dashed lines), one where $\nu_\mathrm{heaviest}$ partially decays and $\nu_\mathrm{middle}$ completely decays (i.e.~$\alpha_\mathrm{heaviest}$, $\alpha_\mathrm{middle} = 10^{-26}$, $10^{-24}~\mathrm{eV^2}$,  dotted lines), and one where $\nu_\mathrm{heaviest}$ completely decays and $\nu_\mathrm{middle}$ partially decays (i.e.~$\alpha_\mathrm{heaviest}$, $\alpha_\mathrm{middle} = 10^{-24}$, $10^{-26}~\mathrm{eV^2}$, dash-dotted lines). The three panels represent the three considered mass configurations: NO SH (leftmost panel), NO QD (middle panel), and IO SH (rightmost panel). The gray band denotes the supernova rate uncertainty and the black curve represents the DSNB in the absence of neutrino decay. The modifications in the DSNB spectrum depend on the specific decay scenario, the decay parameters, and the mass configuration.}

\end{figure}

 Figure~\ref{3nudecaychannelsfig} displays the resultant DSNB spectra in the two scenarios, in red and blue, respectively. 
 
 In the NO SH mass configuration, all $2\nu$ decay channels are degenerate with the no-decay DSNB flux, falling well within the current supernova rate normalization uncertainty band. In addition, both limiting cases we consider involve a channel that increases the integrated flux ($\nu_3 \to \nu_1$ or $\nu_3 \to \nu_2$) and a channel that decreases the integrated flux ($\nu_2 \to \nu_1$). Thus when both channels completely decay, these two effects cancel each other out and the resulting signal is even more degenerate with no decay. We would thus expect to be less sensitive to many $3\nu$ NO SH decay scenarios than their individual $2\nu$ channels. There are significant flux boosts at low energies due to the multiple instances of the SH energy spectra, but not in the detection window. If neutrinos turn out to be in the NO SH mass configuration, it  is thus   difficult to discriminate between no-decay and even complete $3\nu$ decay.

In the NO QD mass configuration, all $2\nu$ channels exhibit the same behavior of increasing the integrated flux and predominantly changing the normalization. Thus, when the channels are combined, the overall behavior is the same, but magnified. There is not a large difference in the signals of the two $3\nu$ limiting cases for the same decay parameters; however, for the same $\alpha_2$ and $\alpha_3$ pair, the $\nu_3 \to \nu_1 + \nu_2 \to \nu_1$ case would boost the DSNB flux more than the $\nu_3 \to \nu_2 + \nu_2 \to \nu_1$ case. These differences grow smaller as the channels decay more completely. In the complete decay regime for both heavy mass states $\nu_3$ and $\nu_2$, the DSNB consists of primarily $\nu_1$ at Earth and results in a $100\%$ $\nu_e$ flux normalization boost compared to the DSNB without decay for both limiting cases.

The IO mass configuration provides the most interesting phenomenology in the $3\nu$ framework, because it contains $2\nu$ channels that both significantly increase and  decrease the DSNB $\nu_e$ flux (see Fig.~\ref{fig:DSNBfluxepartialcompletedecay}). Here, there are marked differences in the phenomenology of the two limiting cases in partial decay regimes.

In the first limiting case (Case I, $\nu_2 \to \nu_3 + \nu_1 \to \nu_3$), all decay products are $\nu_3$, and contribute negligibly to the $\nu_e$ flux. The dominant signature is thus a severe flux reduction.
In the second limiting case (Case II, $\nu_2 \to \nu_1 + \nu_1 \to \nu_3$), all $\nu_2$ must decay through $\nu_1$ before decaying into $\nu_3$. Because $\nu_2 \to \nu_1$ increases the flux and $\nu_1 \to \nu_3$ decreases the flux, the phenomenology associated with this case strongly depends on the relative values of the decay parameters $\alpha_2$ and $\alpha_1$. For example, if $\nu_2$ completely decays while $\nu_1$ only partially decays, the DSNB flux at Earth is primarily comprised of $\nu_1$, and the flux is boosted. If both states completely decay, the DSNB flux at Earth is depleted in a very similar way to the $\nu_2 \to \nu_3 + \nu_1 \to \nu_3$ case, although because of the $\nu_1$ pileup its flux is a bit larger than the first limiting case  for the same complete decay parameters. There are also regimes where the boosting effect of $\nu_2 \to \nu_1$ and the depleting effect of $\nu_1 \to \nu_3$ are roughly equal, and the DSNB signature becomes degenerate with a no-decay scenario.

\begin{figure}[h!]
\centering 
\includegraphics[width=0.99\linewidth]{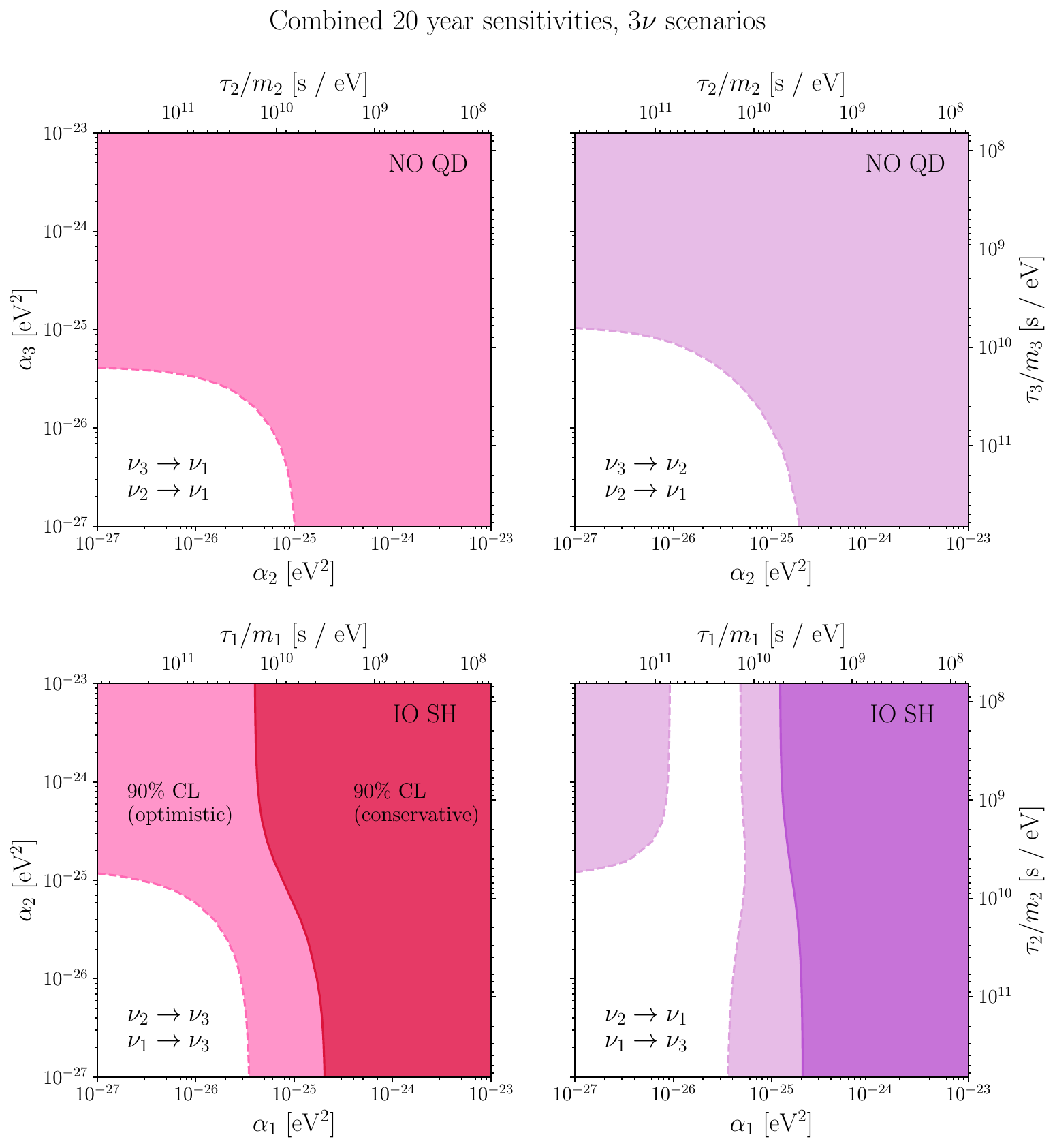}
\caption{\label{fig:3nusensitivities}Expected $90\%$ CL exclusion regions for decay parameter space in representative $3\nu$ decay scenarios assuming a combined Hyper-Kamiokande, JUNO and DUNE  for $20$~years of data taking. \textit{Top panels}: Exclusion regions for two different NO QD decay cases; the left and right panel shows results for the $\nu_3 \to \nu_1 + \nu_2 \to \nu_1$ and for the $\nu_3 \to \nu_2 + \nu_2 \to \nu_1$ scenarios, respectively. We cannot exclude any parameter space at a $90\%$ CL under a conservative detection scenario for these NO QD cases. \textit{Bottom panels:} Exclusion regions for two different IO decay cases; the left and right panels show results for the $\nu_2 \to \nu_3 + \nu_1 \to \nu_3$ and the $\nu_2 \to \nu_1 + \nu_1 \to \nu_3$ scenarios respectively. The $90\%$ CL exclusion regions, assuming a conservative (optimistic) detection scenarios, are plotted in darker (lighter) shades and bordered by solid (dashed) lines. Exclusion regions for the NO SH $3\nu$ decay scenarios are not shown because no parameter space can be excluded at a $90\%$ CL under either a conservative or optimistic detection scenario. 
}
\end{figure}

In Fig.~\ref{fig:3nusensitivities}, we present forecast constraints on these two $3\nu$ decay limiting cases. We expect that, in the limit where one of the mass states barely decays, the signal becomes near-degenerate with a $2\nu$ decay channel, and  the sensitivities should follow a similar pattern of behavior.  If neutrinos follow the NO SH mass configuration, we could not constrain the $3\nu$ neutrino lifetime parameter space at all at a $90\%$ CL in 20 years, even in our optimistic detection scenario. We also find that we are not able to constrain any parameter space at a $90\%$ CL if neutrinos are in a NO QD mass configuration under our conservative detection scenario, but we could achieve stringent constraints in our optimistic scenario. Of note is the IO SH $\nu_2 \to \nu_1 + \nu_1 \to \nu_3$ case, where a portion of partial/full decay parameter space  remains completely degenerate with the null signal; this corresponds to the unconstrained region in the lower right panel of Fig.~\ref{fig:3nusensitivities} for $\alpha_1 \sim 10^{-26}\, {\rm eV}^2$ and $\alpha_2\sim 10^{-24}\, {\rm eV}^2$. 

\bibliographystyle{JHEP}
\bibliography{bibliography.bib}

\end{document}